\documentclass[10pt,twocolumn]{article}

\usepackage{amsmath}
\usepackage{amssymb}

\usepackage{calc} 

\usepackage{multicol}

\usepackage{graphicx}

\usepackage[round,numbers,sort&compress]{natbib} 
\title{Switchable Genetic Oscillator Operating in Quasi-Stable Mode}

\author{{Natalja Strelkowa}\thanks{natalja.strelkowa06@imperial.ac.uk} \, and {Mauricio Barahona}\thanks{m.barahona@imperial.ac.uk, Tel.:~+44 2075945189} \\
Department of Bioengineering \& Institute for Mathematical Sciences\\ Imperial College London \\ South Kensington Campus, London SW7 2AZ, United Kingdom}

\date{}

\begin{document}


\begin{center}
{\LARGE Switchable Genetic Oscillator\\\vspace*{.1in} Operating in Quasi-Stable Mode}
\\ 
\vspace*{.1in}

{{\large Natalja Strelkowa and Mauricio Barahona} \\ \vspace*{.1in} Department of Bioengineering \& Institute for Mathematical Sciences,\\ Imperial College London, \\ South Kensington Campus, London SW7 2AZ, United Kingdom}

\vspace*{.4in}
\end{center} 

\abstract{ Ring topologies of repressing genes have 
qualitatively different long-term dynamics if the number of genes is odd (they oscillate) or even (they exhibit bistability). However, these
attractors may not fully explain the observed behavior in transient and stochastic environments such as the cell. We show here that even repressilators possess quasi-stable, travelling-wave periodic solutions that are reachable, long-lived and robust to parameter changes. These solutions underlie the sustained oscillations observed in even rings in the stochastic regime, even if these circuits are expected to behave as switches. The existence of such solutions can also be exploited for control purposes: operation of the system around the quasi-stable orbit allows us to turn on and off the oscillations reliably and on demand. We illustrate these ideas with a simple protocol based on optical interference that can induce oscillations robustly both in the stochastic and deterministic regimes.

}

\section*{Introduction}

Recent experimental advances in cellular and molecular biology have made
it possible to engineer intricate gene regulatory 
circuits~\cite{Andrianantoandro06}. Inspired in many cases by electronic elements, simple gene networks have been designed to perform reproducible, low-level functions. Some classic examples include the toggle switch~\cite{Gardner00},
the genetic ring oscillator known as the repressilator~\cite{Elowitz00}, or a circuit that can exhibit both oscillatory and switching behavior through the alteration of biochemical interactions~\cite{Atkinson03}. 
Such simple circuits could be potentially interconnected and built-up to form more elaborate `biological devices' with large numbers of components. This trend is facilitated by simulation software containing large number of genes~\cite{Marchisio08} as well as libraries of composable biological parts for experimental realization~\cite{iGemGeneral}. 
Simple synthetic modules can also be integrated into the complex machinery of the cell, as in the oscillator recently implemented in a mammalian cell~\cite{Fussenegger09},
or interfaced with cellular pathways to induce particular responses, as in the construct where the toggle switch was connected to the SOS pathway to induce DNA protection mechanisms in \textit{E. coli} when exposed
to UV light~\cite{Kobayashi04}. Similar principles have been exploited in the rational design of internal negative feedback operated in conjunction with external arabinose-driven positive feedback to produce cell-synchronized oscillations~\cite{Stricker08}.
 
The central role played by oscillations in cellular function has made oscillatory circuits a primary target for the analysis and design of synthetic networks. A particular area of interest is the elucidation of strategies leading to robust timing and sequential activation in the cell. For instance, key stages in developmental biology and in cell differentiation may be controlled by so-called master regulators---a small set of transcription factors sequentially activating and driving several other genes with accurate 
timing~\cite{Borneman06,Liu07,Bondue08}. In addition, studies of both 
natural~\cite{Megerle08,Spencer09} and engineered circuits~\cite{Glick95}
indicate that the correct timing and order of gene activation is a key characteristic
of balanced, optimal cell function, as it reduces the metabolic burdening that ensues from the continuous presence of heterologous substances~\cite{Glieder}. 

In this paper, we consider the dynamics and control of noisy genetic oscillatory
circuits in quasi-stable mode operation. We exemplify our results with
one of the simplest and most widely studied synthetic networks: the $N$-gene ring repressilator (Fig.~1{\it a}). Some natural networks of master 
regulators~\cite{Borneman06} contain such ring structures as subnetworks, making the exploration of their dynamic behavior relevant for both naturally occurring and synthetic systems. The underlying idea is well-known:
when observing the dynamics of systems operating in highly variable environments, such as the cell, it might not be enough to characterize only the long-term attractors of the system since unstable solutions can play a significant role. For instance, quasi-stable transients might be so long-lived as to be the most significant feature of the observed noisy dynamics~\cite{Trefethen}.
Moreover, the presence of noise in nonlinear systems may induce non-stationary dynamics in systems with only fixed point attractors in the deterministic setting~\cite{Suel06} or, conversely, noise may act as a stabilizer of unstable deterministic states~\cite{Turcotte08}.

In the generalized repressilator, results due to Smith~\cite{Smith87,Mueller06} show that rings with an even number of genes 
(e.g., the toggle switch~\cite{Gardner00} with  $N=2$) exhibit multistability and hence behave like switches in the stochastic regime. On the other hand, rings with an odd number of genes (like the standard repressilator~\cite{Elowitz00} with $N=3$) have a globally attracting limit cycle and are therefore oscillators both in the stochastic and deterministic regimes. However, we show here that generalized repressilators possess an intricate structure of unstable periodic orbits that play an important part in their observable noisy and transient dynamics.
In particular, even rings have a quasi-stable limit cycle which, although unstable in terms of linear Floquet stability analysis, has only one unstable direction with a very slow escape rate. This means that trajectories are attracted to the limit cycle from all directions but one, hence leading to long-lived, inducible periodic transients in the deterministic setting and to sustained oscillations in the stochastic system. 
These effects become more pronounced as the number of genes grows. Therefore the finite-time, observable noisy dynamics of an even repressilator ring is not necessarily static (switch-like) but rather exhibits oscillatory characteristics.  

In addition to their effect on the observable dynamics, quasi-stable oscillatory modes can be used
as operating points to control the system around them. The advantage of such a scheme
is that the oscillations can be switched on and off, unlike limit cycle attracting behavior. Operation 
around unstable modes, usually illustrated with the example of the inverted pendulum~\cite{Franklin93}, 
is a classic scenario in control theory for enhanced controllability and speed of response. It has a long 
and successful history of applications in fluid flow control~\cite{Ahuja09} and in the steering of jet 
aircraft~\cite{McRuer04}. Here, we illustrate the application of this concept to gene networks with a simple 
protocol of controlled interference based on an optical mechanism for readout and induction of gene expression. 
Our simulations show that even repressilator rings in quasi-stable operation behave as a robust and on-demand 
switchable oscillator in which genes become upregulated periodically in an ordered sequence according to a 
travelling-wave solution. This characteristic, which is robust at high and low copy numbers, could be used 
for synthetic biological applications such as accurately timed interference with naturally occurring networks.

\section*{Theory}

\subsubsection*{Model equations and stability of fixed points}

The generalized repressilator consists of a ring of $N$ genes in which transcription of each gene is repressed by the product of the preceding gene~(Fig.~1{\it a}). A deterministic model of this circuit is given by the following set of ordinary differential equations:
\begin{eqnarray}
\dot{m}_i&=&\frac{c_1}{1+p_{i-1}^2}-c_2 m_i \nonumber\\
\dot{p}_i&=&c_3m_i-c_4p_i,
\label{eq:system}
\end{eqnarray}
where $p_i$ and $m_i$ describe protein and mRNA concentrations for each gene, respectively~\cite{Elowitz00}.
Here, $i=1,\ldots,N$ with the periodic boundary condition $p_0=p_N$,
and $c_1$ ($c_3$) is the creation rate and $c_2$ ($c_4$) is the degradation rate for the mRNAs (proteins).
The production of mRNA is modeled as a source term that depends nonlinearly on the concentration of the inhibitor protein. Proteins are assumed to be produced at a rate linearly dependent on the amount of the corresponding mRNA. The degradation of mRNA and proteins is assumed to be linearly proportional to their current amount.
The toggle switch~\cite{Gardner00} and the repressilator oscillator~\cite{Elowitz00},
which have both been implemented in \textit{E. coli}, are special cases with $N=2$ and $N=3$, respectively.
Based on analytical results on monotone systems due to Smith~\cite{Smith87,Mueller06},
the stability analysis of this circuit reveals a fundamental difference between rings with odd and even numbers of genes. 
We briefly sketch some of the main results, which are also summarized in Fig.~1{\it b}. 

The stability  analysis characterizes the long-term dynamic behaviors of the deterministic system.  An example of such behavior is given by the fixed points of the system, i.e., the states in which the dynamics is stationary.  The variation of a parameter can produce a change in the stability or the existence of fixed points or other attractors. This is called a bifurcation and it leads to qualitative changes in the long-term behavior of the system. One can find the parameter values at which bifurcations are produced by performing a bifurcation analysis, which can be carried out analytically (in some simple cases) or numerically with the aid of continuation software packages such as AUTO~\cite{AUTO}, which can also track the stability of periodic solutions.

In our system~(\ref{eq:system}), the fixed points, where all derivatives are zero, are found from the condition: 
\begin{equation}
p^*_i \left(1+{p^*_{i-1}}^2\right) = \frac{c_1c_3}{c_2c_4} \equiv c, \quad \forall i.
\label{eq:fixedpoint}
\end{equation}
The parameter $c$ defined in Eq.~(\ref{eq:fixedpoint}) will play the role
of the bifurcation parameter for even rings. 
A positive and real solution is obtained if all proteins have the same concentration:
$p^*_i=p^*_{i+1} = p_m, \forall i$ and
\begin{equation*} 
p_m=\left[\frac{c}{2}+\sqrt{\frac{c^2}{4}+\frac{1}{27}}\right]^{\frac{1}{3}}-\frac{1}{3}
\left[\frac{c}{2}+\sqrt{\frac{c^2}{4}+\frac{1}{27}}\right]^{-\frac{1}{3}}.
\end{equation*}
This solution,
which exists as long as $c$ is positive, is stable for small $c$ and becomes unstable for larger values of $c$ in both odd and even rings.  
   
In the case of even rings, a pitchfork bifurcation takes place at $c=2$ for
all $N$. The two additional stable fixed points arising at that value of the parameter correspond to $p^*_i=p^*_{i+2}\ne p^*_{i+1}, \forall i$, which gives: 
\begin{eqnarray}
p^*_i &=& \frac{c}{2} + \sqrt{\frac{c^2}{4}-1} \quad \equiv \, p_u  \nonumber\\
p^*_{i+1} &=& \frac{c}{2} - \sqrt{\frac{c^2}{4}-1}\quad \equiv \, p_d.
\label{pupd}
\end{eqnarray} 
Note that $p_u \to c-1/c$ and $p_d \to 1/c$ for large $c$.
The new fixed points of the system~(\ref{eq:system}) correspond to two
distinct dimerized states: one in which genes with odd indices are upregulated ($p_u$) while genes with even indices are downregulated ($p_d$); and another symmetric state where the genes with odd and even indices exchange their patterns of regulation. These solutions are equivalent to tiling the ring with copies of the steady state solution of the two-gene ring, and are similar to other dimerized degenerate solutions in classic models of conjugated polymers and spin chains~\cite{Soos07}. Therefore, after the bifurcation, the system is bistable, i.e, it behaves like a switch in the presence of noise.

In the case of odd rings, $p_m$ becomes unstable following a bifurcation that occurs at a value $c(N)$ that approaches $2$ as $N$ grows. However, in this case the bifurcation is Hopf: no additional fixed points appear but rather the bifurcation signals the emergence of a periodic solution. Smith~\cite{Smith87} proved that in monotone systems (i.e., systems in which partial derivatives do not change sign) such as the repressilator, the periodic solution that emerges is a globally attracting stable limit cycle. Therefore odd rings behave as stable oscillators following the Hopf bifurcation.

\subsubsection*{Floquet theory and unstable periodic orbits}
The stability analysis presented above does not provide information about unstable periodic solutions. Although, in principle, unstable periodic orbits are not relevant for the long-term deterministic dynamics, they can be key to the observed dynamics, especially if the orbits involve slow time scales. Such \textit{quasi-stable oscillations} can appear as transients in deterministic simulations and are likely to be observed in the corresponding stochastic simulations. In fact, it was in numerical simulations that we first noticed the relevance of these modes in even repressilator rings.

Floquet theory can be used to find periodic solutions and quantify their linear stability in terms of their Poincar\'e map, i.e., the crossings of the orbit with a (hyper)plane in phase space. Under this analysis, a periodic solution (a closed orbit) becomes a fixed point of the Poincar\'e map and its stability is reformulated as the linear stability of this fixed point.
The eigenvalues of the Poincar\'e map linearized around the fixed point constitute the Floquet multipliers. They indicate how an infinitesimal perturbation around the orbit decays or grows (exponentially).
The periodic solution is linearly stable if all the Floquet multipliers
have magnitudes smaller than unity 
(see the Supplementary Material for references and details on Floquet Theory).
In some cases, a few (possibly only one) Floquet multipliers will be slightly larger than one. We will then have 'quasi-stable' periodic solutions in that it takes a long time to diverge away from them. Quasi-stability in this sense is a local property.  To assess if these solutions will be reachable (and therefore observable in the dynamics), one needs to employ global techniques, e.g. sampling the space of initial conditions. However, Floquet analysis provides an indication of the possibility of observable, yet unstable, periodic solutions. If a periodic orbit has a small number of very weakly unstable directions, it is likely that it could be observed as long-lived periodic transients in the deterministic system and that it could also play a role in the stochastic dynamics. Moreover, such quasi-stable orbits are good candidates for a control mechanism that can induce oscillations on demand, as shown below.

\section*{Methods}
\subsubsection*{Numerical simulations and analysis of the dynamics}
The deterministic system of ODEs~(\ref{eq:system}) was solved numerically with an adaptive fourth-order Runge-Kutta  
integrator~\cite{NumericalRecipesC}, in which the step-size automatically adapts to meet the required accuracy $\epsilon$. 
We have checked that the inducibility, reachability and transient times of the quasi-stable oscillations are not affected by the accuracy of the integrator 
by using the Runge-Kutta integrator with accuracies $\epsilon$ between $10^{-2}$ and $10^{-8}$ (see Supplementary Material for details). In addition, the observability of the unstable orbits was confirmed by using a nonlinear integrator~\cite{SBML-SAT}.

The bifurcation analysis and the calculation of the Floquet multipliers of the unstable periodic orbits were carried out with the numerical 
continuation software AUTO~\cite{AUTO}
(see {\it Supplementary Material}).

Stochastic simulations of the generalized repressilator were performed using the classical Gillespie algorithm~\cite{Gillespie77}. Random numbers and quasi-random numbers for numerical simulations were generated with the GSL Scientific Library~\cite{gsl}.

\subsubsection*{Global robustness analysis and control aspects of quasi-stable oscillations}
 
As part of our numerical evaluation of the generalized repressilator,
we have developed a method to carry out a robustness and reachability analysis of its quasi-stable oscillations. 
This was necessary because available global robustness tools~\cite{SBML-SAT} quantify changes in fixed points induced by parameter variations and are
therefore not directly applicable to oscillations. 
In order to evaluate the global robustness and inducibility of the quasi-stable
oscillations, we attempt to induce sustained oscillations with a predetermined intervention
and quantify changes in the observed response when the model parameters are varied.
The method defines an operating point in parameter space (the reference set  $c_i^*$), based on biologically appropriate estimates, and a hypercube around it to account for biological variability, temperature gradients and other noise. We then sample parameter sets from
the hypercube using Reverse Halton Sequences~\cite{halton60}, quasi-random sequences that have been shown to converge faster than standard Monte Carlo sampling for high dimensional spaces~\cite{Vandewoestyne06}. For each sampled parameter set, we attempt to induce oscillations with the STOP-KICK scenario described below and record if the system evolves
towards sustained oscillations. If oscillations are observed, we calculate
numerically the period of the oscillation and the change in shape
(see supplementary material for details).
Characterizing the change in shape is essential to establish that the oscillation remains detectable and functionally recognizable in the biological system. Note that here we are only concerned with global robustness of the reachability of the solution. A modification of the same
algorithm could be used to study the parameter combinations that contribute most strongly to the sensitivity of the network, a question relevant
for the experimental tuning of the system that is not addressed here.

\section*{Results}
\subsubsection*{Stable and quasi-stable oscillations in the generalized repressilator}

As pointed out in the Theory section, odd repressilator rings are globally attracted to stable limit cycle oscillations for $c > 2$. Numerical simulations show that the period of these solutions increases linearly with the number of genes in the ring (Fig.~2{\it a}). The stability analysis also shows that,
in contrast, even rings only support fixed points as stable solutions. However, direct dynamical simulations of even repressilator rings reveal the existence of long-lived periodic solutions, which are easily reachable, as checked by extensive sampling (not shown) of the space of initial conditions.  The period of these oscillations also increases linearly with the number of genes, albeit with a slope that is approximately half of that in odd rings (Fig.~2{\it a}).

These numerical observations do not pose a contradiction with the stability analysis above: the observed oscillations in even rings are periodic solutions yet unstable. Unstable solutions can be studied using the numerical bifurcation detection software AUTO~\cite{AUTO}, a continuation package that does not rely on dynamical simulations (see Supplementary material). We have used AUTO to find bifurcations in the biologically relevant range of the parameter $c$ and to assess the linear stability of fixed points and periodic solutions---the latter through Floquet analysis. 

The result for even rings is presented in Table~1. In agreement with the analytical results, a pitchfork bifurcation is found numerically as a branching point at $c=2$, above which Hopf bifurcations leading to the appearance of unstable periodic solutions are detected in all rings with more than 4 genes. The Floquet stability analysis indicates that the first unstable periodic orbit to emerge has only one unstable direction, regardless of the number of genes. The only positive Floquet multiplier, which indicates how fast the trajectory diverges away from the orbit, is small and decreases as the length of the ring increases. This is the signature of quasi-stability: if this periodic orbit is reached, it will be long-lived. We have also checked that this solution is reachable through numerical sampling of the space of
initial conditions (see Supplementary material). Such reachable quasi-stable modes affect significantly the observed transient dynamics and also play a central role in stochastic dynamics where unstable solutions are explored under the effect of noise. Both these conditions are relevant for dynamics of genetic circuits inside the cell. 

The existence of quasi-stable modes provides us with the opportunity to design a distinct control strategy. If we control the system to revolve around a quasi-stable mode, the result is an oscillator that can be switched on, kept oscillating and switched off on demand. Below, we introduce a simple
implementation of such a scenario and evaluate its robustness of operation.
Note that an intricate family of unstable periodic orbits with
high symmetry exists both in odd and even rings (see Table~1 and Supplementary Material). However, these additional periodic solutions have several unstable directions that make them essentially unobservable and uncontrollable.

\subsubsection*{Spatio-temporal structure of the periodic solutions}

The spatio-temporal structure of the periodic solutions,  both in the odd
and even cases, corresponds to a travelling-wave solution propagating around the ring. 
The snapshots in Figs.~2{\it b,c} show that this propagation occurs
against the backdrop of the dimerized fixed point solution of the even ring, where all 
odd (even) numbered genes are `up' while the even (odd) numbered genes are `down'~(\ref{pupd}). 
Clearly, a dimerized configuration cannot be accommodated in an odd ring. This leads to a 
kink-like (frustrated) solution where two consecutive genes have similar expression 
levels. This local imbalance of repression induces a dynamical instability that makes the 
kink propagate around the odd ring in a periodic fashion (Fig.~2{\it b}). This spatio-temporal 
structure underlies the limit cycle solution in odd rings. The fact that the period of the limit 
cycle increases (roughly) linearly with the number of genes indicates that the speed of 
propagation of the kink is (roughly) constant.

The quasi-stable periodic solution in even rings can be interpreted under the same prism. 
Figure~2{\it b} shows that it corresponds to \textit{two interacting kinks} propagating around 
the ring at a roughly constant speed with a period that is approximately one half of that of the 
closest odd ring (Fig.~2{\it a}). The instability of this periodic solution has a clear meaning in 
this picture: if the two kinks `collide', they annihilate each other and the system returns to the 
stable fixed point, i.e., the dimerized solution. 
Figure~2{\it b} also shows that each kink has a minimum spatial width that
depends on the parameters of the model. Hence it is more difficult to find these oscillatory solutions 
in rings that are not large enough to fit two such perturbations 
although they can still be observed in smaller, biologically realizable rings (see Supplementary material).
For clarity, we have chosen to illustrate the spatio-temporal structure of the solutions with long rings. 
However, we have checked that the quasi-stable periodic orbits in rings with $N=6, 8, 10$ (not shown) 
maintain the features of the two-kink-structure and operate under the same principles as the long rings shown in Fig.~2{\it b}. 

The spatio-temporal structure of the periodic solutions in repressilator rings shows a strong 
parallelism with similar dynamical solutions observed in classical models of discrete lattices~\cite{braun04}.
The travelling-wave nature of the oscillations could have potential biological applicability if 
one were to use this circuit as a control element for genes that must be activated in a particular 
order and for a pre-defined time interval.

\subsubsection*{Robust induction of quasi-stable oscillations 
in the deterministic regime}
\label{detInduce}

The causal constraints imposed by the travelling-wave structure means that the manipulation of one gene will have a predictable effect on the others. This property can be used to induce and stop oscillations in even rings reliably by activation of one gene for a short time span. We illustrated this simple
scenario in Fig.~3{\it a}. First, the even ring is forced to converge to one of the fixed point solutions with a STOP signal that consists of the external activation of gene $j$ for a time interval longer than the period of the oscillation. This signal is used to `initialize'
the system, suppressing any transient oscillations present in the system. Once the system is at rest, the oscillation can be started with a KICK signal, consisting of
the external activation of gene $j+1$ with a step function of width and amplitude similar to those of the oscillatory pattern. 
Such signals can be imparted non-invasively through an optical mechanism that uses  UV or red light to activate the production of mRNAs of particular genes~\cite{Sato02,Levskaya05}. 

In order to check that the proposed protocol is robust to parameter variations, we have carried out a global robustness analysis as outlined in the Methods section. We construct a hypercube by taking variations of $5\%,10\%$ and $20\%$ around the reference values of the parameters in Eq.~(\ref{eq:system}) and take $10^4$ samples in this hypercube varying all parameters simultaneously. Sampling is performed with quasi-random Reverse Halton Sequences 
for improved convergence~\cite{halton60,Vandewoestyne06}. Figure~3{\it b} shows that the fraction of parameter samples that
lead to oscillations with this protocol converges to 1 for large rings. Our
numerics also show that oscillations can be elicited with significant robustness
in rings with $N >6$.  When oscillations are present, the period shows very
small variation with respect to the reference set, as shown by the coefficient
of variation (Fig.~3{\it b}, inset). We have also quantified the change in the shape of oscillations through a normalized mean square measure and found that the shapes in the perturbed system exhibit very
high similarity ($\approx 99\%$) to the reference set (see Supplementary material). In summary, our global
robustness analysis indicates that in the deterministic regime, long-lived
quasi-stable periodic solutions are reliably inducible in larger rings with a single intervention. 
For moderate size rings, oscillations can still be induced for a large fraction of the parameter hypercube but are shorter
lived. This suggests that repeated interventions could be used in order to keep the ring in the quasi-stable oscillating state. 
A simple control protocol that implements these ideas is proposed in the following section and shown to be applicable for rings as small as $N=6$ operating in the stochastic regime.

\subsubsection*{Stochastic oscillations in even rings and readout-based control}

We have used the standard Gillespie algorithm~\cite{Gillespie77} to study the generalized repressilator in the stochastic regime, i.e., when intrinsic noise is high due to low copy numbers. It is well known that stochastic models of odd rings behave as oscillators and that the travelling wave structure is preserved~\cite{Elowitz00,hembergBJ07}. In the case of even rings, we
have performed long stochastic simulations (not shown) and found bistability and switching events, as expected from the long-term attractors of the underlying deterministic system. Additionally, the simulations show sustained oscillatory behavior, especially in longer rings although they are also observable in rings as small as $N=6$ (see Supplementary material). The oscillations appear in a variety of settings: as transients from a variety of initial conditions; spontaneously emerging from one of the stable points; or associated with switching events. It is also easy to induce such oscillations with localized interventions in particular genes; to sustain them with periodic driving; and to terminate them with a prolonged induction of a gene (as in the STOP signal above). We have examined the structure of these oscillations and they correspond well with the quasi-stable periodic solutions in the deterministic
system: their period is approximately half the period of the closest odd ring (Fig.~2{\it a}) and the spatio-temporal travelling-wave structure is maintained. 

Our numerical simulations confirm the relevance of the underlying quasi-stable oscillations for the observed stochastic dynamics of even rings. Similarly
to the deterministic case, the quasi-stable mode can also be used as a control
operating point such that the system becomes switchable. The robust reachability of this mode allows us to use an extremely simplified feedback mechanism that could be implemented through an optical read-out (GFP, YFP or luciferase protein labeling) and response (on-demand UV or red light gene transcription activation)~\cite{Sato02}. 
The simple control scheme illustrated in Fig.~4{\it a}
uses the optical read-out from two successive proteins in the ring to introduce optical KICK signals that sustain the oscillation based on a threshold rule (Fig.~4{\it b}). The oscillation can be started and terminated using the same optical signals.
Although we have chosen to illustrate the possible implementation of the control scheme with light sensitive inducers, it is worth remarking that any suitable mechanism capable of precisely-timed gene induction with good spatial resolution at the cell population could be used.  
A potential advantage of this switchable mode of operation is the economical and targeted use of the transcriptional resources without overburdening the cell with unnecessary mRNA production~\cite{Glieder}.

\section*{Discussion}

In this work, we have studied how the presence of quasi-stable periodic solutions
affects the observable dynamics of even repressilator rings. Previously, even rings have been thought of as switches due to the fact that they only support fixed
point solutions.  However, our bifurcation analysis reveals the existence of a set of unstable orbits, some of which have slow timescales associated with them. These quasi-stable periodic solutions are both reachable and long-lived, thus playing a role in the observed dynamics, both transient and stochastic. This suggests that oscillatory behavior might be more widespread than expected in genetic models since it could feature in systems that possess only static attractors.

The presence of quasi-stable solutions provides us with
the possibility of designing control protocols that operate the system around such modes so that the oscillations can be turned on and off reliably. Our
numerics indicate that a robust mechanism could be implemented based on appropriate
optical feedback to switch the system between stable fixed points and quasi-stable oscillations. 
Although the proposed pared-down control scheme is only intended to provide
an illustration of the potential implementation and  its performance could be improved with the use of optimized strategies for stochastic and robust 
control that take into account specific details of the experimental setup, it is worth discussing some of its limitations.  The challenge for the dynamical control
algorithm is to deliver the optical interference signal necessary for  the induction of gene expression for a short time period, to a particular spatial area of the
cell population, and with a well-controlled delay following the fluorescent signal of the proteins. The proposed scheme shows both enough spatial and time resolution to address individual cells in a population with well-defined pulses~\cite{Sato02}.  The scheme would need to rely on proper calibration of life-times of fluorescent proteins affected by phototoxic and photo-bleaching effects, as reviewed in detail by \citet{Bennett09}. Finally, since the delay between the actual protein concentration and the corresponding fluorescent signal introduced by the maturation time 
($\sim$ 2-8 min) is short compared to the period of the oscillation ($\sim$ 130 min), this would be acceptable for the control scheme.  

A synthetic circuit operating under such principles could be interfaced with a naturally occurring network to induce an intrinsic interference
that is interruptible on-demand. The switchability of this regulatory element can help avoid the appearance of adverse cumulative effects. 
The NF$\kappa$B pathway is an example where such a
regulator could provide controlled activation over short time intervals as
an alternative to conventional knock-downs and other functional interventions that modify the balance of important proteins for the cell
cycle~\cite{nfkbReview,nfkbCurrentOp}. 
The underlying travelling wave structure of the observed periodic solutions could also be potentially useful for design purposes. It allows for coordinated intervention when the timing and order of activation of different pathways is crucial. Examples of cellular networks, e.g. in developmental biology, indicate that timed patterns of sequential activation are at the heart of the function of families of master regulators~\cite{ Ashall09,Holtzendorff04,Liu07,Spencer09} and, in the case of  the vertebrate segmentation clock~\cite{Yun-Jin00, Pourquie03}, the associated oscillations are well-defined but do not survive in the long-term. The importance of heterogeneously timed gene induction has also been highlighted
in a model of Arabinose uptake in \textit{E. coli}~\cite{Megerle08}.
Experiments with genetically engineered yeast have also shown that pulsed activation of chaperons followed by pulsed activation of the associated heterologous proteins is more efficient at maximizing the production of particular metabolites~\cite{Glieder}. These applications hint at potential uses for circuits that can produce sequential patterns of activation on demand, such as the even repressilator studied here, which interact with other cellular pathways via intrinsic proteins, thus avoiding the timed delivery of external agents through the cell membrane.

The design of control strategies for the operation of systems around an inherently unstable state 
has a long history in other disciplines (e.g., flight and fluid control) since it affords enhanced 
responsiveness and controllability with relative ease and simplicity of design~\cite{Franklin93}. 
This strategy differs fundamentally from the biochemical alteration of the 
network topology proposed by~\citet{Atkinson03} 
based on a smaller gene circuit but with a complex regulatory scheme involving
promoter and repressor sites regulating one gene. 
The molecular kinetics of such regulators are less well understood than those with single regulatory sites 
due to unavoidable cross-talk and compound logic. The ring topology relies on simple regulation to provide a sequence of causal signals but at the expense of involving a larger number of genes.

The present scheme is also in contrast with previously engineered gene circuits, such as odd repressilators, which possess 
globally attracting limit cycles leading to behavior that is robust yet not controllable. Quasi-stable 
operation, on the other hand, is robustly switchable. 
The switchability of the oscillator coupled with dynamic control that affords good spatial resolution 
could be used to elicit localized oscillations in cell populations as an aid to examine mechanisms 
of cell synchronization.
It is an open area of current research to elucidate the role of a design
concept based on control around unstable behavior, similar to the inverted pendulum in classic control theory, 
to further our understanding of cell strategies and its potential
use in the design of synthetic topologies that can interfere with naturally occurring pathways.

\section*{Acknowledgements}
N.S. gratefully acknowledges support from the Wellcome Trust through grant 080711/Z/06/Z. We thank Manos Drakakis, 
Karen Polizzi, Andreas Ipsen and Tomasz Leja for inspiring discussions and suggestions.
%
{\small
\bibliography{references1}

\begin{thebibliography}{43}
\providecommand{\url}[1]{\texttt{#1}}
\providecommand{\urlprefix}{ }

\bibitem[Andrianantoandro et~al.(2006)Andrianantoandro, Basu, Karig, and
  Weiss]{Andrianantoandro06}
Andrianantoandro, E., S.~Basu, D.~K. Karig, and R.~Weiss, 2006.
\newblock Synthetic biology: new engineering rules for an emerging discipline.
\newblock \emph{Mol. Sys. Bio.} 2:2006.0028.

\bibitem[Gardner et~al.(2000)Gardner, Cantor, and Collins]{Gardner00}
Gardner, T., C.~R. Cantor, and J.~J. Collins, 2000.
\newblock Construction of a genetic toggle switch in Escherichia coli.
\newblock \emph{Nature} 403:339--342.

\bibitem[Elowitz and Leibler(2000)]{Elowitz00}
Elowitz, M.~B., and S.~Leibler, 2000.
\newblock A synthetic oscillatory network of transcriptional regulators.
\newblock \emph{Nature} 403:335--338.

\bibitem[Atkinson et~al.(2003)Atkinson, Savageau, Myers, and Ninfa]{Atkinson03}
Atkinson, M.~R., M.~A. Savageau, J.~T. Myers, and A.~J. Ninfa, 2003.
\newblock Development of Genetic Circuitry Exhibiting Toggle Switch or
  Oscillatory Behavior in Escherichia coli.
\newblock \emph{Cell} 113:597--607.

\bibitem[Marchisio and Stelling(2008)]{Marchisio08}
Marchisio, M., and J.~Stelling, 2008.
\newblock {Computational design of synthetic gene circuits with composable
  parts}.
\newblock \emph{Bioinformatics} 24:1903--1910.

\bibitem[iGe()]{iGemGeneral}
MIT Registry. http://partsregistry.org. Accessed June 6, 2009 .

\bibitem[Tigges et~al.(2009)Tigges, Marquez-Lago, Stelling, and
  Fussenegger]{Fussenegger09}
Tigges, M., T.~T. Marquez-Lago, J.~Stelling, and M.~Fussenegger, 2009.
\newblock A tunable synthetic mammalian oscillator.
\newblock \emph{Nature} 457:309--312.

\bibitem[Kobayashi et~al.(2004)Kobayashi, Kaern, Araki, K., Gardner, Cantor,
  and Collins]{Kobayashi04}
Kobayashi, H., M.~Kaern, M.~Araki, C.~K., T.~S. Gardner, C.~R. Cantor, and
  J.~J. Collins, 2004.
\newblock Programmable cells: interfacing natural and engineered gene networks.
\newblock \emph{Proc. Natl. Acad. Sci. USA} 101:8414--8419.

\bibitem[Stricker et~al.(2008)Stricker, Cookson, Bennett, Mather, Tsimring, and
  Hasty]{Stricker08}
Stricker, J., S.~Cookson, M.~R. Bennett, W.~H. Mather, L.~S. Tsimring, and
  J.~Hasty, 2008.
\newblock A fast, robust and tunable synthetic gene oscillator.
\newblock \emph{Nature} 456:516--519.

\bibitem[Borneman et~al.(2006)Borneman, Leigh-Bell, Yu, Bertone, Gerstein, and
  Snyder]{Borneman06}
Borneman, A.~R., J.~A. Leigh-Bell, H.~Yu, P.~Bertone, M.~Gerstein, and
  M.~Snyder, 2006.
\newblock Target hub proteins serve as master regulators of development in
  yeast.
\newblock \emph{Genes Dev.} 20:435--448.

\bibitem[Liu et~al.(2007)Liu, Asakura, Inoue, Nakamura, Sano, Niu, Chen,
  Schwartz, and Schneider]{Liu07}
Liu, Y., M.~Asakura, H.~Inoue, T.~Nakamura, M.~Sano, Z.~Niu, M.~Chen, R.~J.
  Schwartz, and M.~D. Schneider, 2007.
\newblock {Sox17 is essential for the specification of cardiac mesoderm in
  embryonic stem cells}.
\newblock \emph{Proc. Natl. Acad. Sci. USA} 104:3859--3864.

\bibitem[Bondue et~al.(2008)Bondue, Lapouge, Paulissen, Semeraro, Iacovino,
  Kyba, and Blanpain]{Bondue08}
Bondue, A., G.~Lapouge, C.~Paulissen, C.~Semeraro, M.~Iacovino, M.~Kyba, and
  C.~Blanpain, 2008.
\newblock Mesp1 Acts as a Master Regulator of Multipotent Cardiovascular
  Progenitor Specification.
\newblock \emph{Cell Stem Cell} 3:69--84.

\bibitem[Megerle et~al.(2008)Megerle, Fritz, Gerland, Jung, and
  R\"adler]{Megerle08}
Megerle, J.~A., G.~Fritz, U.~Gerland, K.~Jung, and J.~O. R\"adler, 2008.
\newblock Timing and Dynamics of Single Cell Gene Expression in the Arabinose
  Utilization System.
\newblock \emph{Biophys. J} 95:2103--2115.

\bibitem[Spencer et~al.(2009)Spencer, Gaudet, Albeck, Burke, and
  Sorger]{Spencer09}
Spencer, S.~L., S.~Gaudet, J.~G. Albeck, J.~M. Burke, and P.~K. Sorger, 2009.
\newblock Non-genetic origins of cell-to-cell variability in TRAIL-induced
  apoptosis.
\newblock \emph{Nature} 459:428--432.

\bibitem[Glick(1995)]{Glick95}
Glick, B.~R., 1995.
\newblock Metabolic load and heterologous gene expression.
\newblock \emph{Biotechnology Advances} 13:247 -- 261.

\bibitem[Glieder(2009)]{Glieder}
Glieder, A., 2009.
\newblock Technische Universit\"at Graz. Private Communication .

\bibitem[Trefethen and Embree(2005)]{Trefethen}
Trefethen, L.~N., and M.~Embree, 2005.
\newblock Spectra and pseudospectra.
\newblock Princeton University Press, Princeton, NJ, USA.

\bibitem[S\"uel et~al.(2006)S\"uel, Garcia-Ojalvo, Liberman, and
  Elowitz]{Suel06}
S\"uel, G.~M., J.~Garcia-Ojalvo, L.~M. Liberman, and M.~B. Elowitz, 2006.
\newblock An excitable gene regulatory circuit induces transient cellular
  differentiation.
\newblock \emph{Nature} 440:545--550.

\bibitem[Turcotte et~al.(2008)Turcotte, Garcia-Ojalvo, and S\"uel.]{Turcotte08}
Turcotte, M., J.~Garcia-Ojalvo, and G.~M. S\"uel., 2008.
\newblock {A genetic timer through noise-induced stabilization of an unstable
  state}.
\newblock \emph{Proc. Natl. Acad. Sci. USA} 105:15732--15737.

\bibitem[Smith(1987)]{Smith87}
Smith, H., 1987.
\newblock Oscillations and multiple steady states in a cyclic gene model with
  repression.
\newblock \emph{J. Math. Biol.} 25:169--190.

\bibitem[M\"uller et~al.(2006)M\"uller, Hofbauer, Endler, Flamm, Widder, and
  Schuster]{Mueller06}
M\"uller, S., J.~Hofbauer, L.~Endler, C.~Flamm, S.~Widder, and P.~Schuster,
  2006.
\newblock A generalized model of the repressilator.
\newblock \emph{J. Math. Biol.} 53:905--937.

\bibitem[Franklin et~al.(1993)Franklin, Emami-Naeini, and Powell]{Franklin93}
Franklin, G.~F., A.~Emami-Naeini, and J.~D. Powell, 1993.
\newblock Feedback Control of Dynamic Systems.
\newblock Addison-Wesley Longman Publishing Co., Inc., Boston, MA, USA.

\bibitem[Ahuja and Rowley(2009)]{Ahuja09}
Ahuja, S., and C.~W. Rowley, 2009.
\newblock Feedback control of unstable steady states of flow past a flat plate
  using reduced-order estimators arXiv.org:0902.1207.

\bibitem[McRuer and Graham(2004)]{McRuer04}
McRuer, D., and D.~Graham, 2004.
\newblock Flight Control Century: Triumphs of the Systems Approach.
\newblock \emph{J. of Guidance, Control, and Dynamics} 27:161--173.

\bibitem[Doedel(2007)]{AUTO}
Doedel, E., 2007.
\newblock AUTO 07. Software for Continuation and Bifurcation Problems in
  Ordinary Differential Equations .

\bibitem[Soos(2007)]{Soos07}
Soos, Z., 2007.
\newblock Identification of dimerization phase transitions driven by Peierls
  and other mechanisms.
\newblock \emph{Chemical Physics Letters} 440:87--91.

\bibitem[Press et~al.(1992)Press, Teukolsky, Vetterling, and
  Flannery]{NumericalRecipesC}
Press, W.~H., S.~A. Teukolsky, W.~T. Vetterling, and B.~P. Flannery, 1992.
\newblock Numerical recipes in C: the art of scientiﬁc computing.
\newblock Cambridge University Press, 2 edition.

\bibitem[Zi et~al.(2008)Zi, Zheng, Rundell, and Klipp]{SBML-SAT}
Zi, Z., Y.~Zheng, A.~E. Rundell, and E.~Klipp, 2008.
\newblock SBML-SAT: a systems biology markup language (SBML) based sensitivity
  analysis tool.
\newblock \emph{BMC Bioinformatics} 9.

\bibitem[Gillespie(1977)]{Gillespie77}
Gillespie, D.~T., 1977.
\newblock Exact Stochastic Simulation of Coupled Chemical Reactions.
\newblock \emph{J. Chem. Phys.} 8:2340--2361.

\bibitem[Galassi et~al.()Galassi, Davies, Theiler, Gough, Jungman, Alken,
  Booth, and Ross]{gsl}
Galassi, M., J.~Davies, J.~Theiler, B.~Gough, G.~Jungman, P.~Alken, M.~Booth,
  and F.~Ross.
\newblock GNU Scientific Library Reference Manual.
\newblock 3 (v1.12) edition.

\bibitem[Halton(1960)]{halton60}
Halton, J., 1960.
\newblock On the efficiency of certain quasi-random sequences of points in
  evaluating multi-dimensional integrals.
\newblock \emph{Numer. Math. 2} 84--90.

\bibitem[Vandewoestyne and Cools(2006)]{Vandewoestyne06}
Vandewoestyne, B., and R.~Cools, 2006.
\newblock Good permutations for deterministic scrambled Halton sequences in
  terms of L2-discrepancy.
\newblock \emph{J. Comput. Appl. Math.} 189:341--361.

\bibitem[Braun and Kivshar(2004)]{braun04}
Braun, O., and Y.~Kivshar, 2004.
\newblock The Frenkel-Kontorova Model. Concepts, Methods, and Applications.
\newblock Springer, illustrated edition.

\bibitem[Shimizu-Sato et~al.(2002)Shimizu-Sato, Huq, Tepperman, and
  Quail]{Sato02}
Shimizu-Sato, S., E.~Huq, J.~M. Tepperman, and P.~H. Quail, 2002.
\newblock A light-switchable gene promoter system.
\newblock \emph{Nat. Biotech} 20:1041--1044.

\bibitem[Levskaya et~al.(2005)Levskaya, Chevalier, Tabor, Simpson, Lavery,
  Levy, Davidson, Scouras, Ellington, Marcotte, and Voigt]{Levskaya05}
Levskaya, A., A.~A. Chevalier, J.~J. Tabor, Z.~B. Simpson, L.~A. Lavery,
  M.~Levy, E.~A. Davidson, A.~Scouras, A.~D. Ellington, E.~M. Marcotte, and
  C.~A. Voigt, 2005.
\newblock Synthetic biology: engineering Escherichia coli to see light.
\newblock \emph{Nature} 438:441--2.

\bibitem[Hemberg and Barahona(2007)]{hembergBJ07}
Hemberg, M., and M.~Barahona, 2007.
\newblock Perfect Sampling of the Master Equation for Gene Regulatory Networks.
\newblock \emph{Biophys. J.} 93:401--410.

\bibitem[Bennett and Hasty(2009)]{Bennett09}
Bennett, M.~R., and J.~Hasty, 2009.
\newblock Microfluidic devices for measuring gene network dynamics in single
  cells.
\newblock \emph{Nat Rev Genet} 10:628--638.

\bibitem[Karin and Lin(2002)]{nfkbReview}
Karin, M., and A.~Lin, 2002.
\newblock NF-$\kappa$B at the crossroads of life and death.
\newblock \emph{Nat Immunol} 3:221--227.

\bibitem[Naugler and Karin(2008)]{nfkbCurrentOp}
Naugler, W.~E., and M.~Karin, 2008.
\newblock NF-$\kappa$B and cancer - identifying targets and mechanisms.
\newblock \emph{Current Opinion in Genetics \& Development} 18:19--26.

\bibitem[Yun-Jin et~al.(2000)Yun-Jin, Aerne, Smithers, Haddon, Ish-Horowicz,
  and Lewis]{Yun-Jin00}
Yun-Jin, B.~L. Aerne, L.~Smithers, C.~Haddon, D.~Ish-Horowicz, and J.~Lewis,
  2000.
\newblock Notch signalling and the synchronization of the somite segmentation
  clock.
\newblock \emph{Nature} 408:475--479.

\bibitem[Pourquie(2003)]{Pourquie03}
Pourquie, O., 2003.
\newblock {The Segmentation Clock: Converting Embryonic Time into Spatial
  Pattern}.
\newblock \emph{Science} 301:328--330.

\bibitem[Ashall et~al.(2009)Ashall, Horton, Nelson, Paszek, Harper, Sillitoe,
  Ryan, Spiller, Unitt, Broomhead, Kell, Rand, See, and White]{Ashall09}
Ashall, L., C.~A. Horton, D.~E. Nelson, P.~Paszek, C.~V. Harper, K.~Sillitoe,
  S.~Ryan, D.~G. Spiller, J.~F. Unitt, D.~S. Broomhead, D.~B. Kell, D.~A. Rand,
  V.~See, and M.~R.~H. White, 2009.
\newblock {Pulsatile Stimulation Determines Timing and Specificity of
  NF-{kappa}B-Dependent Transcription}.
\newblock \emph{Science} 324:242--246.

\bibitem[Holtzendorff et~al.(2004)Holtzendorff, Hung, Brende, Reisenauer,
  Viollier, McAdams, and Shapiro]{Holtzendorff04}
Holtzendorff, J., D.~Hung, P.~Brende, A.~Reisenauer, P.~H. Viollier, H.~H.
  McAdams, and L.~Shapiro, 2004.
\newblock {Oscillating Global Regulators Control the Genetic Circuit Driving a
  Bacterial Cell Cycle}.
\newblock \emph{Science} 304:983--987.

\end{thebibliography}


{\small 
\begin{thebibliography}{}
\bibitem{Kampen07} N. G. van Kampen (2007)
{\it Stochastic Processes in Physics and Chemistry.}
3-rd Edition. (Elsevier, Amsterdam)
\bibitem{Vandewoestyne06} B. Vandewoestyne and R. Cools (2006)
Good permutations for deterministic scrambled Halton sequences in terms of L2-discrepancy
{\it J. Comput. Appl. Math.} 189:341--361
\bibitem{Strogatz} S. H. Strogatz (1994)
{\it Nonlinear Dynamics and Chaos.}
(Westview Press)

\bibitem{Doedel} E Doedel (2007) 
AUTO 07. Software for Continuation and Bifurcation Problems in Ordinary Differential Equations
\end{thebibliography}
}
}
\clearpage
\pagestyle{empty}
\parbox[c]{\textwidth}{
\parbox[c]{0.49\textwidth}{{\small 
\begin{tabular}{|c|c|c|c|c|c|}
\hline
N & Bifur-& c &Stable/all& Max. & Period\\
  & cation&   &directions& Floquet & (min)\\
\hline
2& BP & 2.00 & - & -&-\\
\hline
4& BP & 2.00 & - & -&-\\
\hline
6& BP & 2.00 & - & -&-\\
& HB & 4.68 & 11/12 & 6.2&132\\
\hline
8& BP & 2.00 &  - & -&-\\
& HB & 3.00 & 15/16 & 3.9&186\\
\hline
10& BP & 2.00 & - & -&-\\
& HB & 2.56 & 19/20 & 3.0&239\\
& HB & 9.62 & 17/20 & 9.0&104\\
\hline
12& BP & 2.00 & - & -&-\\
& HB & 2.36 & 23/24 & 2.5&291\\
& HB & 4.68 & 21/24 & 6.2&132\\
\hline
14& BP & 2.00 & - & -&-\\
& HB & 2.24 & 27/28 & 2.2 & 342\\
& HB & 3.51 & 25/28 & 4.8 & 160\\
& HB & 16.9 & 23/28 & 10.5 & 94\\
\hline
16& BP & 2.00 & - & -&-\\
& HB & 2.18 & 31/32 & 2.0 &393\\
& HB & 3.00 & 29/32 & 3.9 & 186 \\
& HB & 6.91 & 27/32 & 7.8 & 113\\
\hline
18& BP & 2.00 & - & -&-\\
& HB & 2.15 & 35/36 & 1.8 & 444\\
& HB & 2.71 & 33/36 & 3.4 & 213\\
& HB & 4.71 & 31/36 & 6.2 & 132\\
& HB & 27.33 & 29/36 & 11.5 & 89\\
\hline
\end{tabular}
 }
}
\parbox[c]{0.49\textwidth}{
\includegraphics*[width=2.05in]{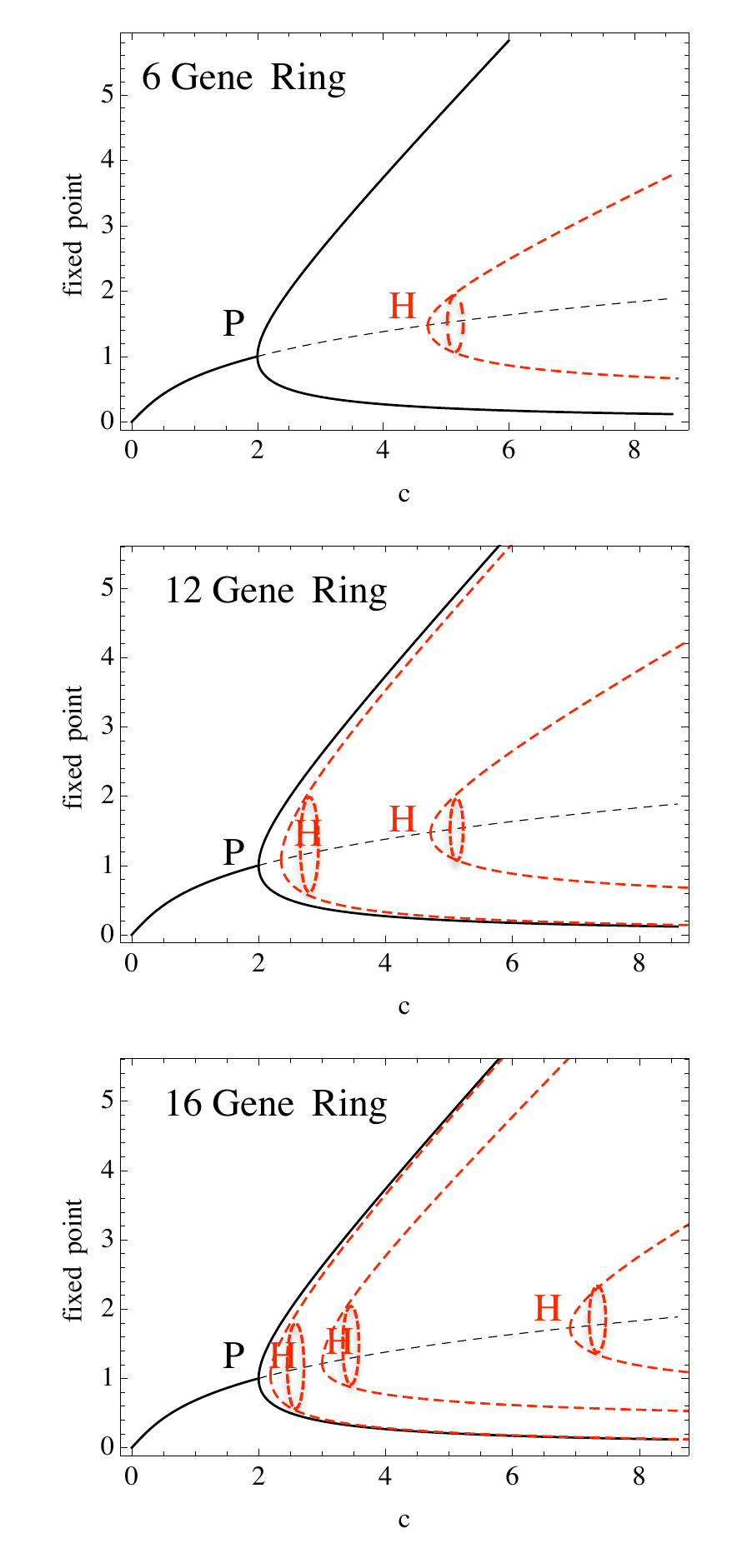}
}
}
\parbox[c]{\textwidth}{\small{\textbf{Bifurcation analysis and unstable periodic solutions 
of repressilator rings with even number of genes.}  We use the continuation package 
AUTO~\cite{AUTO} to obtain the bifurcations of rings of size 
$N$~(\ref{eq:system}). The parameter $c$, defined in Eq.~(\ref{eq:fixedpoint}), is swept 
in the biologically relevant range $c \in [0.001,30]$ by changing
$c_1$ with $c_2=0.12$, $c_3=0.16$ and $c_4=0.06$ constant. In agreement with 
analytical calculations, a branching point (P), corresponding to a pitchfork bifurcation, 
is found at $c=2$. A series of Hopf bifurcations (HB) linked to the emergence of unstable 
periodic solutions are found subsequently. Floquet analysis indicates that the first unstable 
orbit to emerge has only one unstable direction, regardless of the dimension of the system, 
and that the maximal Floquet multiplier decreases with increasing $N$. Hence this periodic 
solution is quasi-stable: if it is reached, the divergence away from it is slow, and gets 
slower for longer rings. Other unstable orbits are present but their high instability makes 
them irrelevant to the observed dynamics. A similar structure of unstable orbits exists in 
odd rings (see Supplementary Material).
The figure on the right shows the bifurcation diagrams for even rings of length $N=6, 12, 16$. The unstable
periodic orbits, shown in red dashed lines, emerge through Hopf bifurcations.
}}

\clearpage

\section*{Figure Legends}
\begin{figure*}
   \begin{center}
      \includegraphics*[width=4.25in]{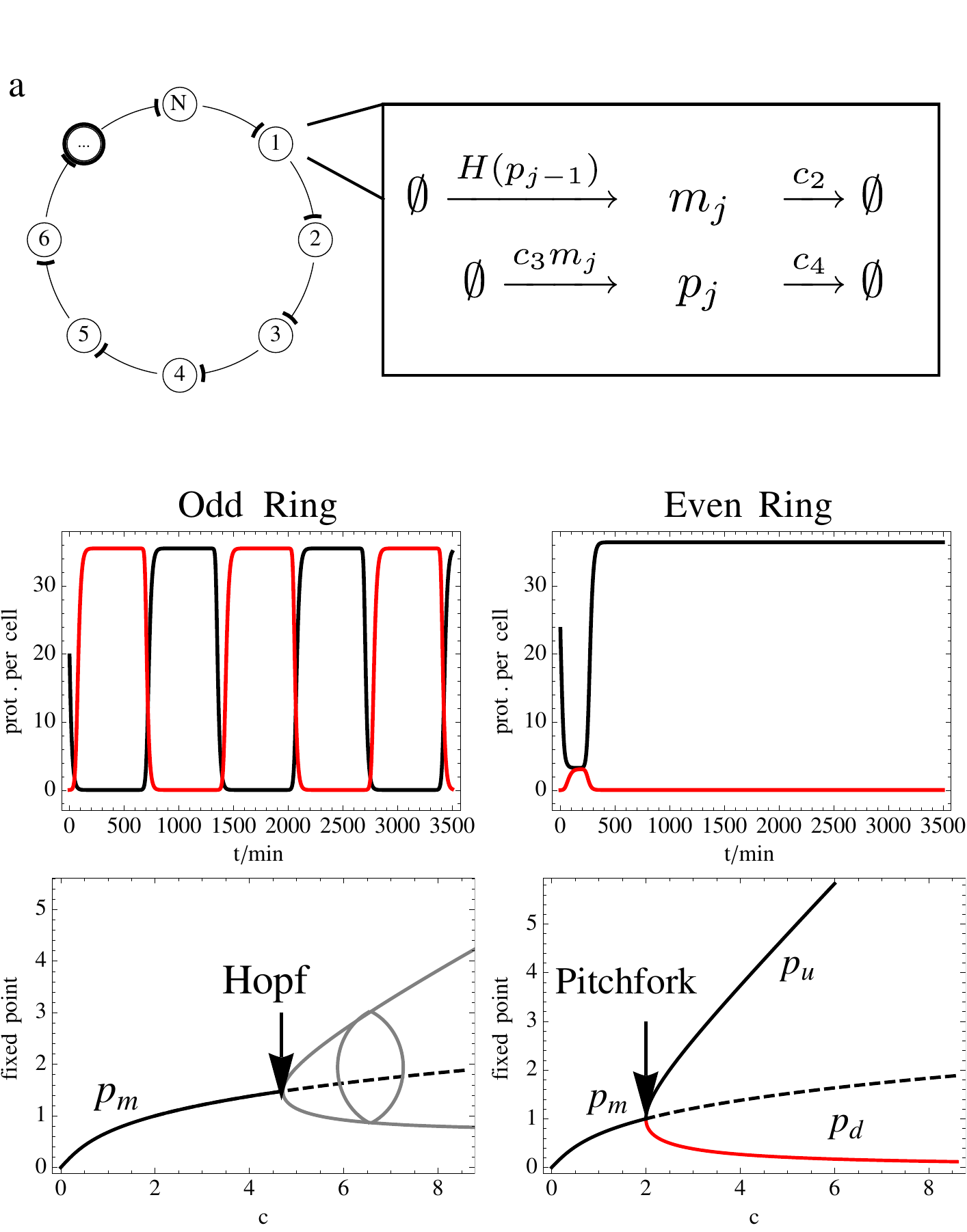}
      \caption{
      }
      \label{fig:result_fig}
   \end{center}
\end{figure*}

\subsubsection*{Figure~\ref{fig:result_fig}. Attractors of the generalized repressilator model}
{\small {\it (a)} Topology of the generalized repressilator: $N$ genes in a cycle where each gene is 
repressed by the protein product of the preceding gene. Also shown is the reaction scheme 
underlying the dynamical system~(\ref{eq:system}) with production and degradation
terms for the mRNA ($m_j$) and protein ($p_j$) of each gene. The repression of the production of 
mRNA is modelled by a Hill-type term $H(p_{j-1})$. {\it (b)} Typical time traces of the long-term 
deterministic dynamics of an odd ring and an even ring above the bifurcation point $c=2$: odd 
rings converge to a globally attracting periodic solution while even rings converge to fixed points. 
The time traces shown correspond to $N=23$ and $N=22$. {\it (c)} Stability of the fixed points of 
the system as a function of the bifurcation parameter~$c$.  Even rings undergo a pitchfork bifurcation 
at $c=2$, leading to the emergence of two stable fixed points.  Odd rings undergo a Hopf bifurcation 
leading to the emergence of a limit cycle. The critical parameter for the Hopf bifurcation depends on 
$N$ but tends to $c=2$ as $N$ grows (see Supplementary material). }

\subsubsection*{Figure~\ref{fig:result_fig_2}. Periodic solutions and travelling waves in the generalized repressilator model}
{\small
{\it (a)} The period of the limit cycle of the deterministic model of odd rings 
(solid line) increases linearly with the length of the ring. Simulations
of the stochastic version of this system using the Gillespie algorithm show
that the period (shown as circles) follows the same trend, although they are 
slightly larger. The period of the quasi-stable solutions found in even rings 
(deterministic and stochastic) increases linearly with the number of genes but 
with a slope that is half that of the odd rings. The inset shows
representative time traces of the periodic solutions in odd rings (stable)
and in even rings (quasi-stable). 
{\it (b)} Time snapshot of the spatial distribution 
of the concentrations of two successive proteins concentrations for the periodic 
solution in the odd ring with $N=23$. The solution has a
travelling-wave structure with a kink-like perturbation propagating around
the ring, 
indicated by the arrow in the bottom figure.  The bottom figure represents the minimum distance $|\Delta p_j|=\min( |p_u-p_j| , |p_d-p_j| )$ between the traveling wave solution and the dimerized solution with an alternating pattern of protein expression given by  $p_u$ and $p_d$.  The distance becomes large around the kink in the traveling wave solution.
{\it (c)} Same as (b) for the 
quasi-stable periodic solution of the even ring with $N=22$. In this case, the 
travelling wave solution has two kinks that propagate around the ring, as indicated by the arrows. 
}

\subsubsection*{Figure~\ref{fig:result_fig_3}. Robust induction of oscillations
in even rings in the deterministic regime}
{\small {\it (a)} The quasi-stable periodic solution in even repressilator rings can be induced 
with a simple sequence of signals. First, apply a STOP signal to gene $j$ to force the system 
to approach a fixed point solution. Second, apply a KICK signal to gene $j+1$ to drive the 
ring into oscillation. The oscillation can be terminated at will by applying another STOP signal. 
The signals can be implemented via on-demand UV or red light gene transcription activation~\cite{Sato02}.
This STOP-KICK-STOP protocol is shown here for a ring with $N=18$ with parameters
$c_1=2.6$, $c_2=0.12$, $c_3 = 0.2$, $c_4 = 0.06$.
{\it (b)} Global robustness analysis of the inducibility of the quasi-stable oscillations. The STOP-KICK 
scenario is applied to $10^4$ random combinations of parameters $c_i$ for each even ring of length 
$N$ and we record the proportion of parameter sets that lead to five oscillations. The parameters are sampled with reverse Halton sequences from a hypercube with $5\% (\blacksquare), 10\% (\bullet)$ and $20\% (\blacklozenge)$ variation  around the reference set. Quasi-stationary oscillations are robustly induced for $N \geq 10$, while smaller rings can be kept in the oscillating state applying repeated
interventions in a simple control protocol as shown in Fig.~\ref{fig:result_fig_4}. (Inset) The oscillations are robust in shape (not shown) and in period to changes in the parameters. The relative variability (coefficient of variation) of the period of the induced oscillations is small and decreases with the length of the ring.}

\subsubsection*{Figure~\ref{fig:result_fig_4}. Stochastic oscillations in
even rings and readout-based control.}
{\small 
{\it (a)} Illustration of the readout-based control scheme for a ring of 6 genes. Two proteins of the ring are
read out with fluorescent tags. This readout is then compared to a reference defined according to the 
oscillating behavior of the ring, with similar period and a shift between consecutive genes. 
The reference comparison is threshold-based and leads to an ON-OFF (1-0) control for 
the KICK signals. These can be implemented with light responsive genes promoters. 
In the numerical simulations shown in (b), the KICK signals are indicated with the red markings in the upper panels.
{\it (b)} A simple readout-based control reliably switches on the oscillations, sustains them and switches them 
off. The control mechanism functions by monitoring two successive proteins in the ring. Whenever each of 
them falls below a threshold, a KICK signal for the corresponding protein is given. 
These threshold-based KICK signals are indicated with red and magenta markings in the upper panels.
The oscillation can be terminated 
with a STOP signal as in the deterministic state.  The optical read-out can be based on GFP or YFP protein labeling while the response can 
be implemented with on-demand UV or red light that enhances the production of the corresponding mRNAs~\cite{Sato02}. 
The figure shows the application of this mechanism to a ring with $N=10$. The stochastic time traces correspond to the 
protein expression of proteins $p_j$ with $j=1, 3, 5 , 7 , 9$ and the corresponding control (top) in response to proteins $p_1$ and $p_2$
( trace not shown).
The right
figure is a magnification of the dashed square inside the main figure.  We have also checked that this control protocol is applicable for rings with as low as $N=6$ genes (not shown).}

\clearpage


\clearpage
\begin{figure*}
\centerline{\includegraphics*[width=5.25in]{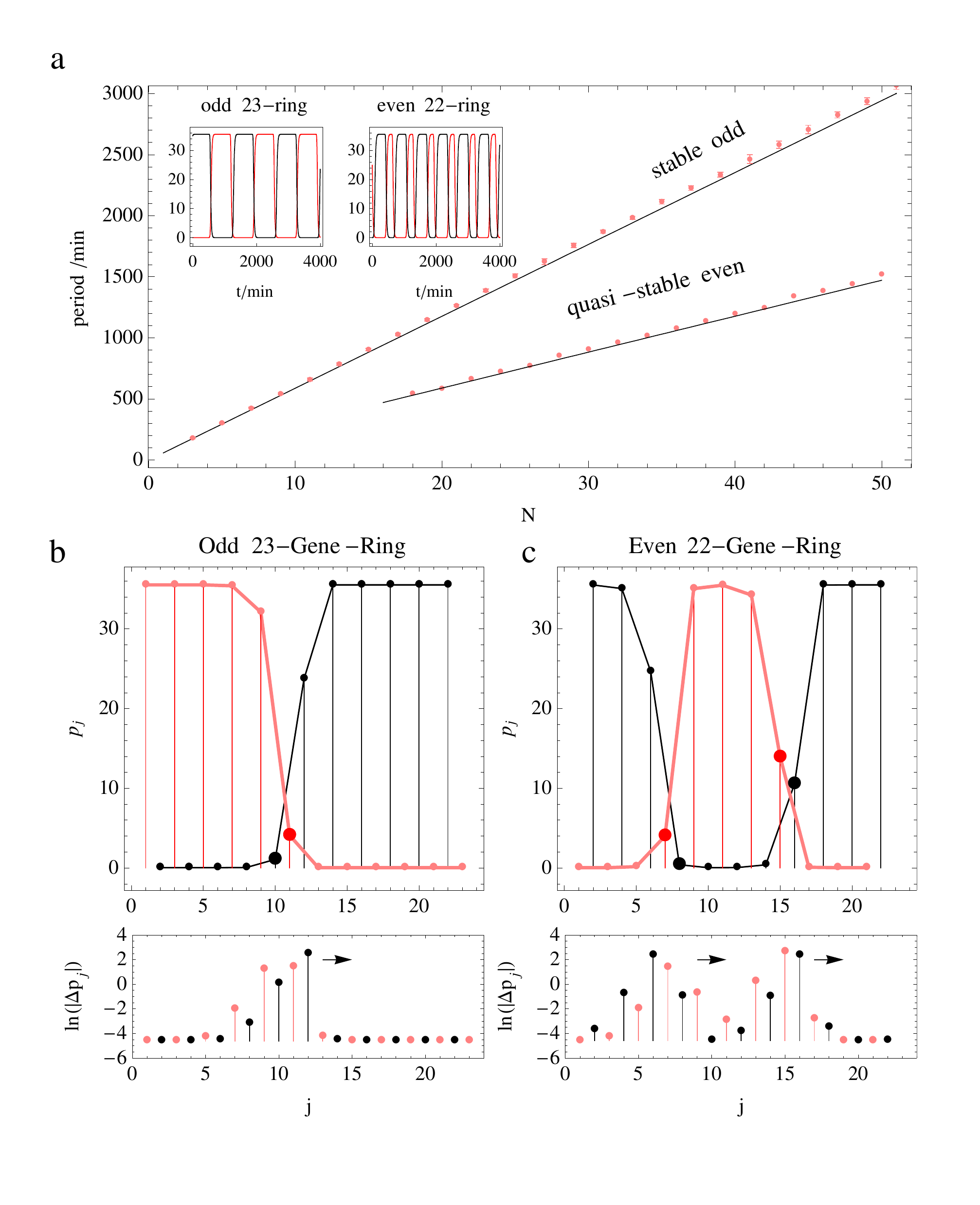}}
      \caption{
}
      \label{fig:result_fig_2}
\end{figure*}

\clearpage
\begin{figure*}
   \begin{center}
      \includegraphics*[width=7in]{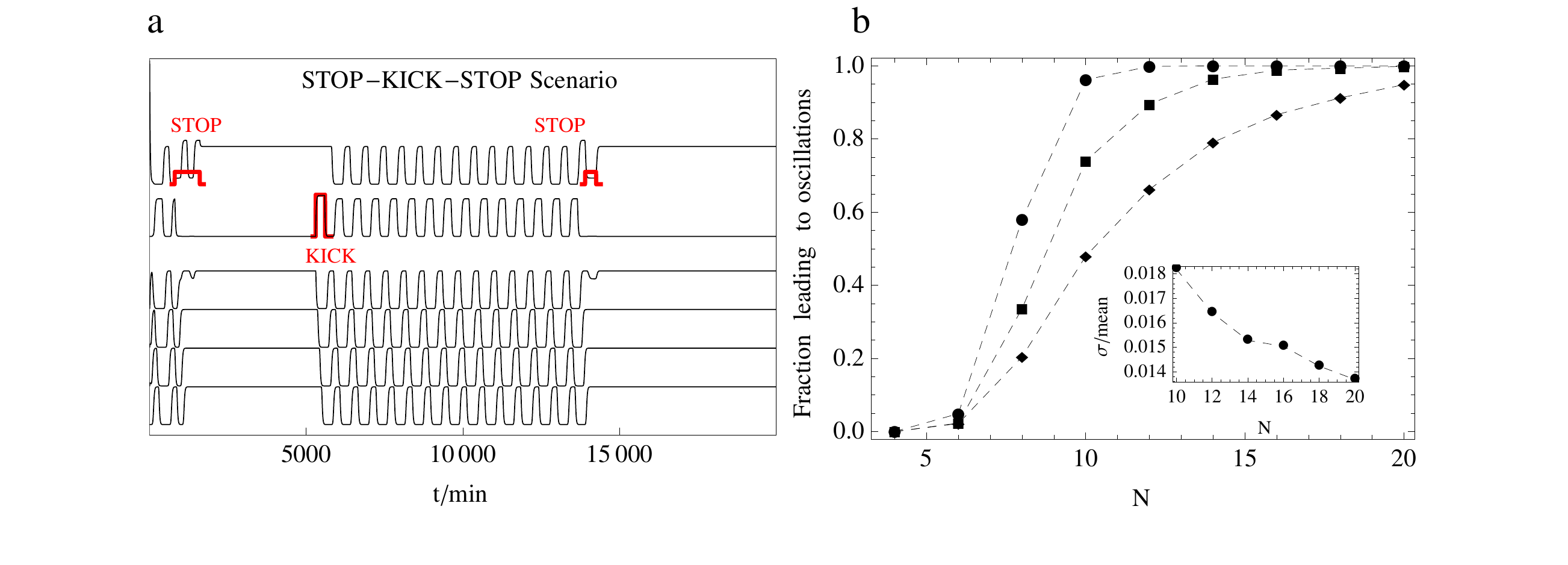}
      \caption{ 
}
      \label{fig:result_fig_3}
   \end{center}
\end{figure*}

\clearpage
\begin{figure*}
   \begin{center}
      \includegraphics*[width=6.5in]{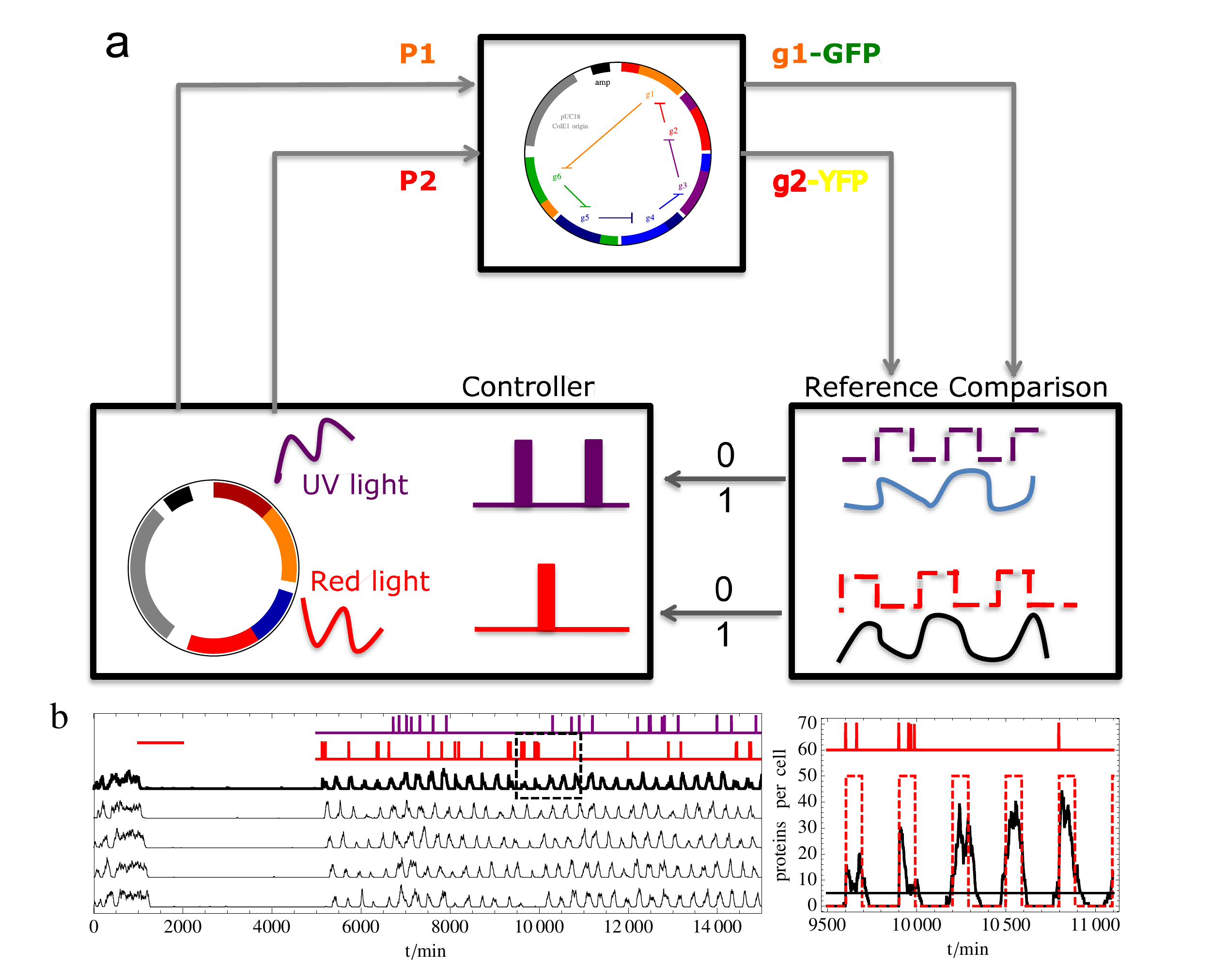}
      \caption{
      }
      \label{fig:result_fig_4}
   \end{center}
\end{figure*}

\clearpage
\onecolumn
\begin{center}
{\Large \textbf{Supplementary Material}}
\end{center}

\section{Stability analysis and bifurcations in odd and even rings}


\parbox[c]{\textwidth}{
\parbox[c]{0.6\textwidth}{
{\small 
\begin{tabular}[h]{|c|c|c|c|c|c|}
\hline
N & Bifur-& c &Stable/all& Max. & Period\\
  & cation&   &directions& Floquet & (min)\\
\hline
\hline
3& HB & 4.689 & 6/6 & 1 &132\\
\hline
5 & HB & 2.556 & 10/10 & 1&239\\
\hline
7& HB & 2.244 & 14/14 & 1 & 342\\
& HB & 16.91 & 12/14 & 8.0 & 94\\
\hline
9& HB & 2.156 & 18/18 & 1 & 444\\
& HB & 4.711 & 16/18 & 5.0 & 132\\
\hline
11& HB & 2.089 & 22/22 & 1 & 545\\
& HB & 3.311 & 20/22 & 3.75 & 169\\
& HB & 42.84 & 18/22 & 11.0 & 86\\
\hline
13& HB & 2.067 & 26/26 & 1 & 645\\
& HB & 2.800 & 24/26 & 3.05 & 204\\
& HB & 7.778 & 22/26 & 7.55 & 110\\
\hline
15& HB & 2.044 & 30/30 & 1 & 746\\
& HB & 2.533 & 28/30 & 2.63 & 239\\
& HB & 4.711 & 26/30 & 5.76 & 132\\
\hline
17& HB & 2.044 & 34/34 & 1 & 847\\
& HB & 2.400 & 32/34 & 2.35 & 274\\
& HB & 3.667 & 30/34 & 4.68 & 154\\
& HB & 11.75 & 28/34 & 9.13 & 100\\
\hline
\end{tabular} }}
\parbox[c]{0.4\textwidth}{\includegraphics*[width=2.05in]{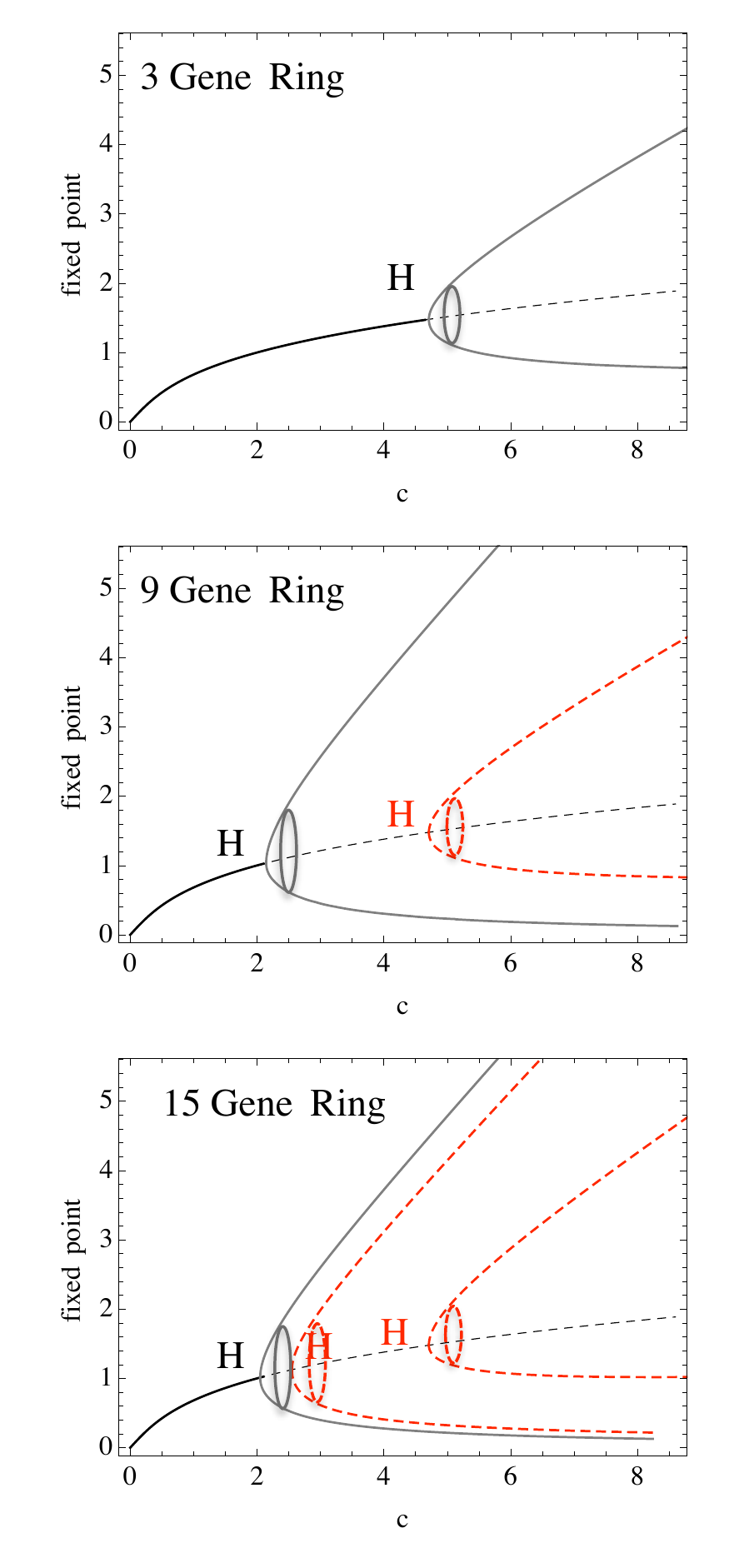}}
}
\parbox[c]{\textwidth}{\small{\textbf{Table SM1: Bifurcation analysis  of repressilator rings with odd number of genes and corresponding bifurcation diagrams.}}}

\clearpage

\begin{figure*}[h!]
   \begin{center}
      \includegraphics*[width=5.25in]{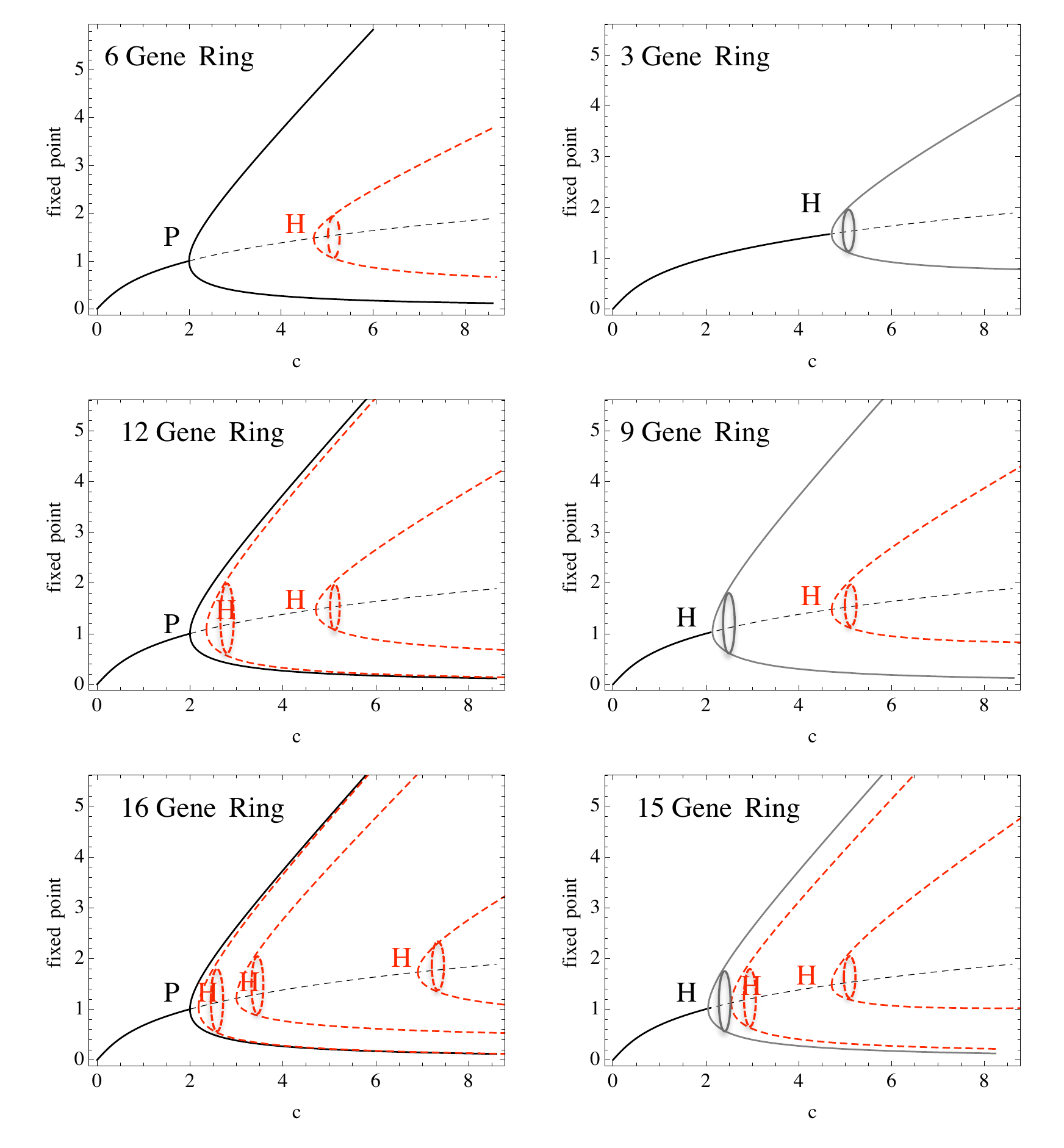}
      \caption{Bifurcation diagrams for odd and even case. The unstable periodic orbits are shown in  red dashed lines. As the number of genes in the ring increases the first unstable orbit moves towards
stable structures (stable fixed points for the even rings and stable limit cycles for the odd rings. This suggests that the first unstable orbits may become more observable in larger rings. The sampling of initial condition space shown below confirms this hint.)}
      \label{bifurDiags}
   \end{center}
\end{figure*}

Table SM1 and Figure~\ref{bifurDiags} summarize the result of our analysis using AUTO to obtain the bifurcations of odd rings of size $N$. 
As in the main text, the parameter $c$ is swept in the biologically relevant range $c \in [0.001,30]$.
In agreement with analytical calculations summarized in the main text, a series of Hopf bifurcations (HB) are found. The first one leads to a stable limit cycle,
whose maximum Floquet multiplier is 1 corresponding to the trivial direction
along the orbit. The subsequent bifurcations give birth to unstable periodic orbits with an increasing number of unstable directions (always even). 
These unstable orbits are essentially irrelevant to the observed dynamics because they are highly unstable and not reachable. The
relationships between the periods of oscillations in rings of different lengths
hints at a strong symmetry in the spatio-temporal structure of these solutions. We summarize the findings in Table SM1 and Figure~\ref{bifurDiags}.

\section{Basic Floquet Theory and Quasi-Stable Oscillating Orbits under Noise}
\begin{figure*}[h!]
   \begin{center}
      \includegraphics*[width=3.25in]{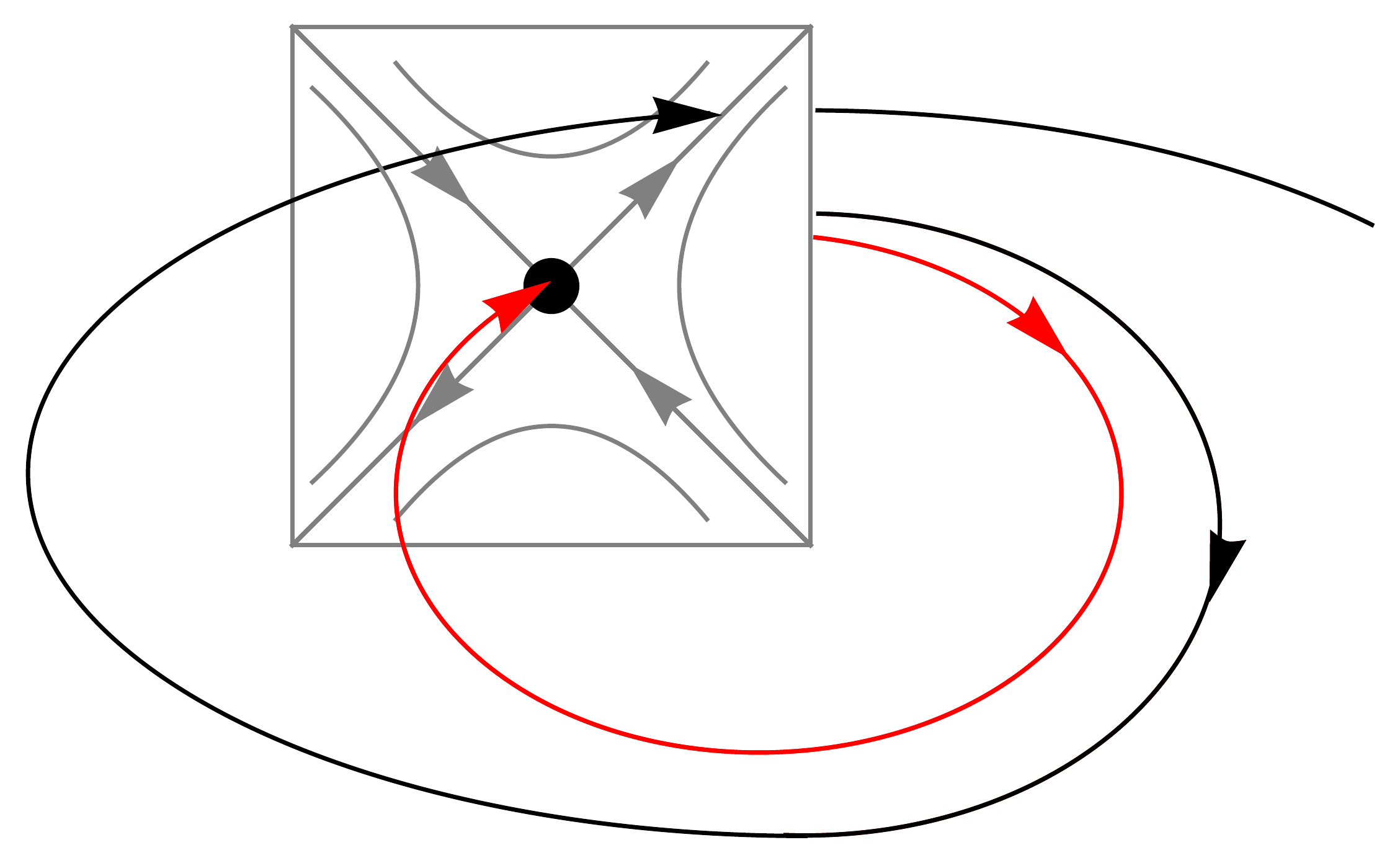}
      \caption{Poincare map with quasi-stable fixed point (black) and possible trajectories. The periodic orbit (red trajectory) is linearly stable along the left diagonal, where the arrows are pointing towards the fixed point and unstable along the right diagonal, where the arrows are pointing away from the fixed point. The escape scenario from unstable periodic orbit due to pertubations is illustrated with black trajectory.}
      \label{poincare}
   \end{center}
\end{figure*}

As pointed out in the main text we have observed long lasting oscillations in the even rings randomly sampling the initial condition space.
The observed oscillating modes can be assessed regarding their stability with the Floquet Multipliers theory \cite{Strogatz}. Here, the question about stability of the orbit can be reformulated to the question for the stability of the corresponding fixed point $x^*$ in the Poincare map. If $v_0$ is an infinitesimal perturbation such that $x^* + v_0$ is in $S$, then after the first return to $S$
\begin{equation}
x^* + v_1 = P(x^* + v_0) \approx P(x^*) + [DP(x^*)] v_0
\end{equation}
The Eigenvalues $\lambda_j$ of the linearized Poincare map $[DP(x^*)]$ obtained this way are nontrivial Floquet multipliers. The closed orbit is linearly stable if and only if $|\lambda_j|<1$ for all $j$. We illustrate the case where the closed orbit has unstable directions and how the system may escape from it in Figure~\ref{poincare} above. The Poincare section (shown in gray) has stable directions (left diagonal) and unstable directions (right diagonal). The system may escape the closed orbit (black trajectory) in a directed manner once perturbed along the unstable direction.

Since it is a biological system it is important to show that the quasi-stable orbits are observable under noise effects. We can simulate small biological noise effects using less accurate integration with larger maximal error $\epsilon$ per step (see  "Methods" section in the main text). This is in addition to the robustness analysis under parameter variations illustrated in Figure~{\bf SS1}. It turns out that the oscillations do not become less robust and last as long as with the accurate integration.

\parbox[c]{\textwidth}{
\parbox[c]{0.65\textwidth}{
\hspace{-0.1cm}
\includegraphics*[width=3.5in]{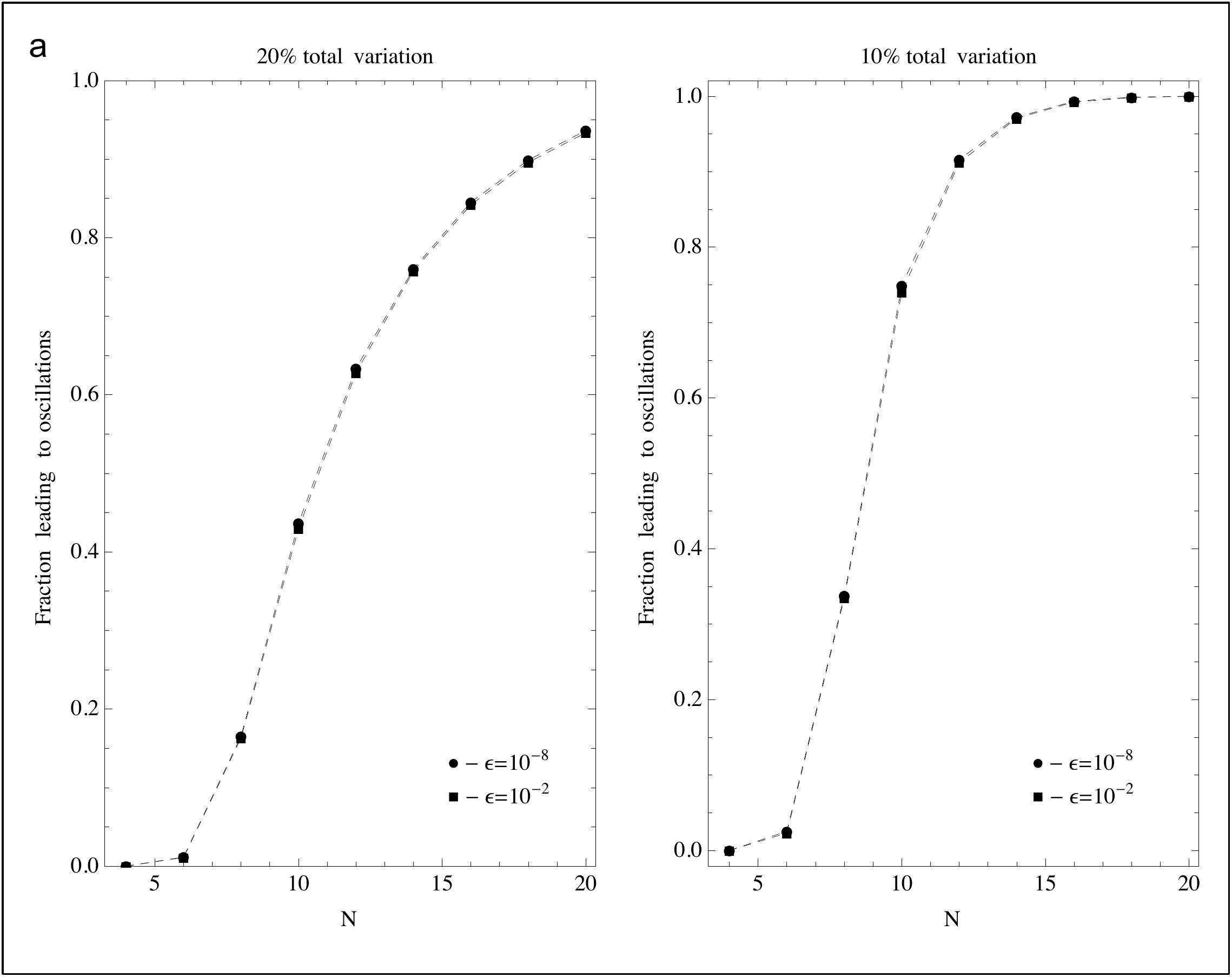}
}
\parbox[c]{0.30\textwidth}{
\includegraphics*[width=1.75in]{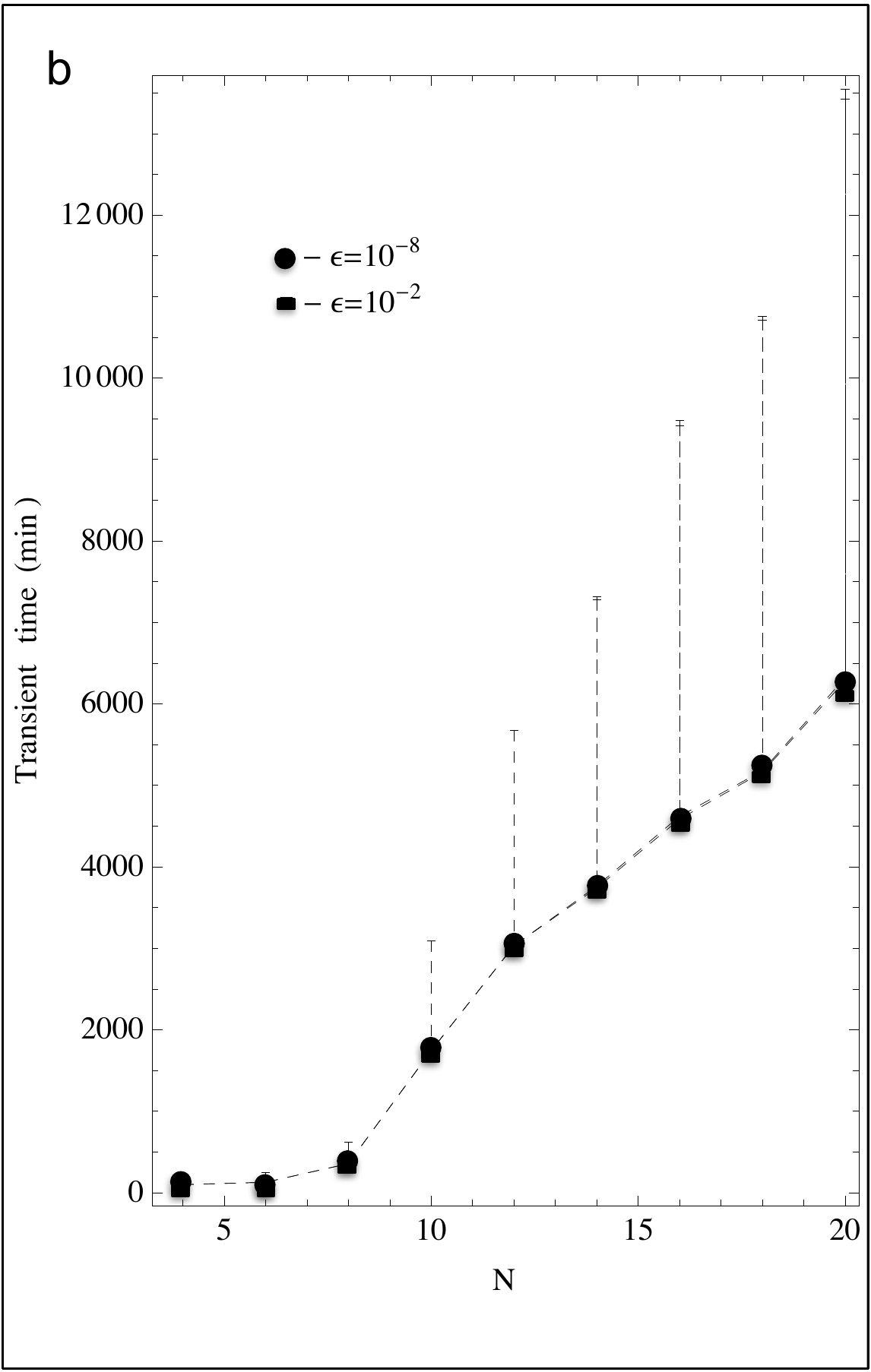}
}
}
\parbox[c]{\textwidth}{
 Figure SS1: {\bf a} Robustness of STOP-KICK-STOP scenario as illustrated in the main text Figure 3a. The accuracy control parameter $\epsilon$ for circles is $10^{-2}$, whereas in the main text the parameter was $10^{-8}$. Both curves are for the same total variation. Comparing the result to the main text, we see that less accurate integration does not change the robustness of the STOP-KICK-STOP scenario significantly. Less accurate integration is a way to simulate the biological noise and these quasi-stable oscillations do not become significantly less robust under the noise simulated with less accurate integration as shown here and also simulated with Gillespie algorithm as shown later in the control scheme in the main text. (See the paragraph below for the intuitive explanation for this phenomenon.) {\bf b} Different accuracy also does not significantly affect the length of transient oscillations, which die off within a certain period of time (for instance within 50 oscillations). The figure illustrates the mean lengths of oscillations, which occur if starting from the randomly chosen initial condition and inspect the region for 50 oscillations. The error bars are first and third quartiles.
\vspace{2cm}
}

We try to give an intuitive explanation for the observed effect. Errors resulting from less accurate integration affect all variables describing protein and mRNA concentrations. It has the effect that the system is perturbed in all directions due to this noise. However, the generalized repressilator system can escape the oscillating orbit through directed movement along one out of many possible directions (see Figure~\ref{poincare} above  and Table~1 in the main text). If the system is perturbed along other directions, it is pulled back to the limit cycle, because all other directions are attactive. Therefore random perturbations do not neccessarily kill the transient oscillations. On the contrary, random perturbations may interfer with the directed escape out of the limit cycle and this might even stabilize the oscillations. In fact in the following section we show fluctuations due to internal noise stabilize the oscillations.

\section{Sampling of Initial Condition Space and Spontanious Transient Oscillations in the Stochastic Setting}
\begin{figure*}[h!]
   \begin{center}
      \includegraphics*[width=3in]{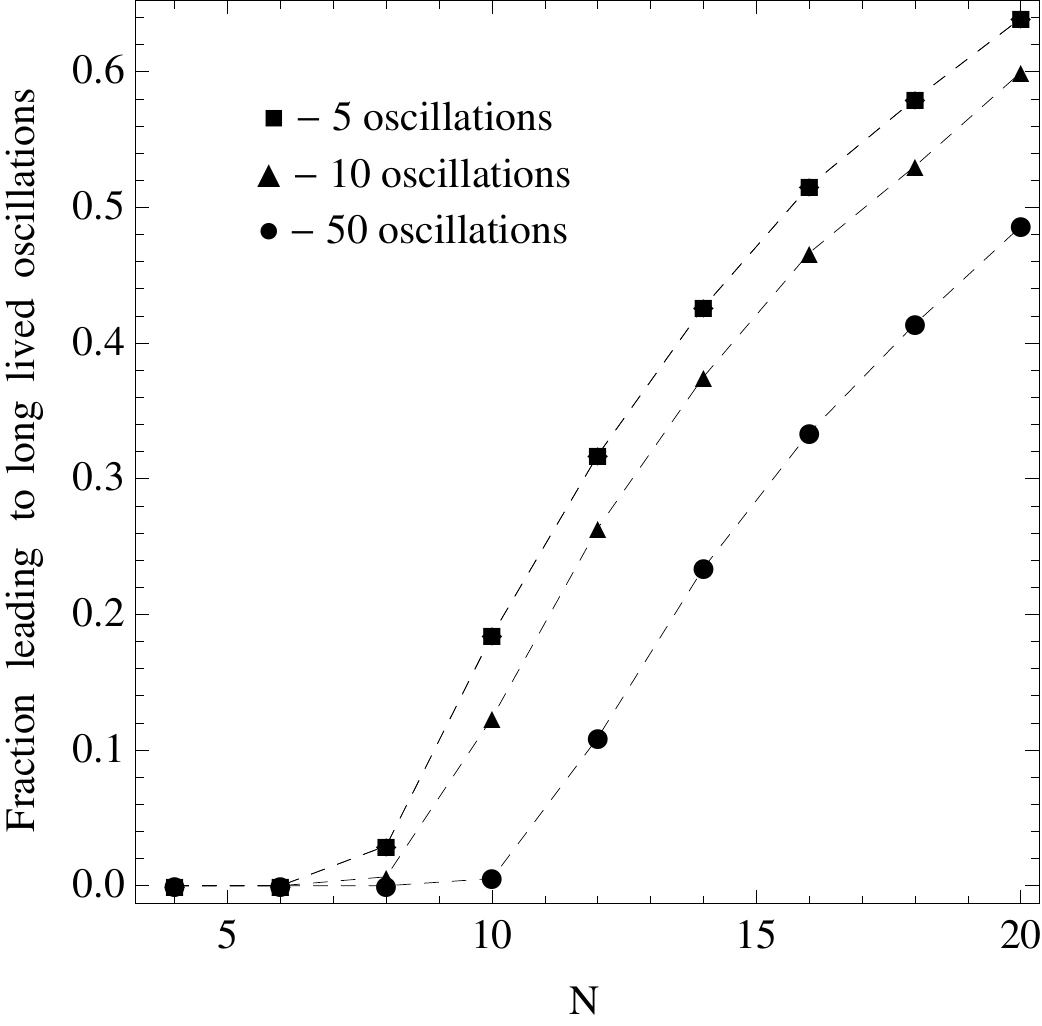}
      \caption{Sampling of the initial condition space for the deterministic simulations. The initial condition hyper cube defined as 0 to maximal amplitude 
(Note: the maximal amplitude of the proteins is not the same as the maximal amplitude for the mRNAs) has been 
sampled $10^4$ times using pseudo random numbers (reverse Halton sequences).  }
      \label{initSpaceSampling}
   \end{center}
\end{figure*}

The bifurcation analysis revealed the existence of quasi-stable periodic solutions in the generalized repressilator model. 
However, it is not clear if these solutions are observable under random sampling of the initial condition space. Here we
show for the deterministic setting how often on average a randomly sampled initial condition can lead to at least 5,10, 
and as many as 50 oscillations Fig.~\ref{initSpaceSampling}.

In the stochastic setting the noise due to small copy numbers does not destroy the oscillations in the even rings. On the contrary, 
this is also how we have discovered these modes. The quasi-stable structure identified with the deterministic bifurcation theory 
becomes even more prevalent under internal noise. So, in this case noise does actually excite the oscillations. The following parameters 
have been used $c_1=1.6, c_2=0.12, c_3=2.6,c_4=0.06$. The picture shows that in the stochastic regime the system spontaneously goes 
into the oscillating state more frequently than the equivalent deterministic description. The initial conditions for the stochastic simulation 
are chosen to be the same as in the deterministic case and 10 runs for each starting value has been collected.

\begin{figure*}[h!]
   \begin{center}
      \includegraphics*[width=5in]{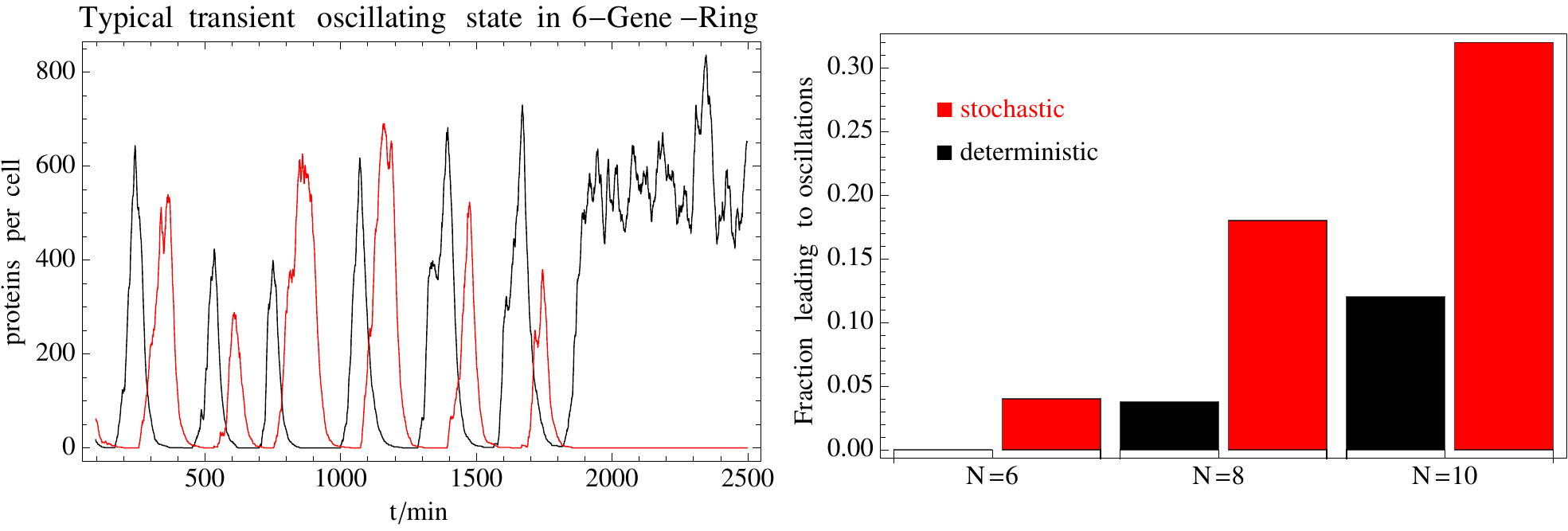}
      \caption{Transient oscilations are more prevalent in the stochastic setting. For the same initial conditions and the same constants 
the system goes more frequently into the oscillating state than in the deterministic setting.}
      \label{stochRes}
   \end{center}
\end{figure*}

\section{Robustness Analysis and Shape Similarity}
We perform global robustness analysis, which means all parameters are varied at the same time within specified multidimensional
hypercube. If we want to obtain a global robustness of let say 10 gene ring, then the dimension of the according parameter space 
would be 40, there are 4 constants for each gene ($c_1,c_2,c_3,c_4$) and these are varied independently of each other. This procedure is fundamentally different to
standard robustness  assessments, where only one parameter is gradually varied at the same time.

The parameter hypercube is created as $ 5\%, 10\%, 20\%$ variation around the reference state, where all parameters $c_i$ are the same for all the genes.
We sample this parameter hypercube in an efficient manner using pseudo-random numbers created with reverse Halton sequences.
This sampling of large dimensional space with pseudo-random numbers was shown to be more efficient than the classical MC \cite{Vandewoestyne06}.
After selecting a set of parameters in this way, we evolve the system in time first applying the STOP signal, which puts it into stationary up/down solution.
Then we apply a KICK scenario and record if the oscillations have persisted for at least 5 periods. The requirement here is that all genes have shown a non decaying
amplitude. In this way we obtain $10^4$ samples for each gene number and plot the fraction of the samples leading to the oscillations on the y-Axes.

In addition to the 0,1 decision on if or not the fraction oscillates we quantify the change in the shape of the oscillation. This is independent of the variation in period
and amplitude. The measure $ssh$ would give large dissimilarity scores if the oscillations would become spiky, the gene being in the up state is much smaller than being
in the down state or vise versa. It would give intermediate scores if instead of being rather quadratic, it becomes like a Gaussian shape. 
Mathematically, we first scale the time for the reference shape and for the perturbed system shape to 1. Through the scaling we get rid of the variations in the period lengths.
Then we apply the standard normalized $L_2$-norm,
\begin{equation}
ssh = < \pi_{ref}(t) \pi_{pert}(t) >/<\pi_{ref}(t)>/<\pi_{pert}(t)>
\end{equation}
where the normalization scales away the differences in the amplitude. This measure assesses shape changes
for moderate parameter perturbations and gives an additional information to the variation in the oscillation amplitude and period length.
\begin{figure*}[h!]
   \begin{center}
      \includegraphics*[width=6in]{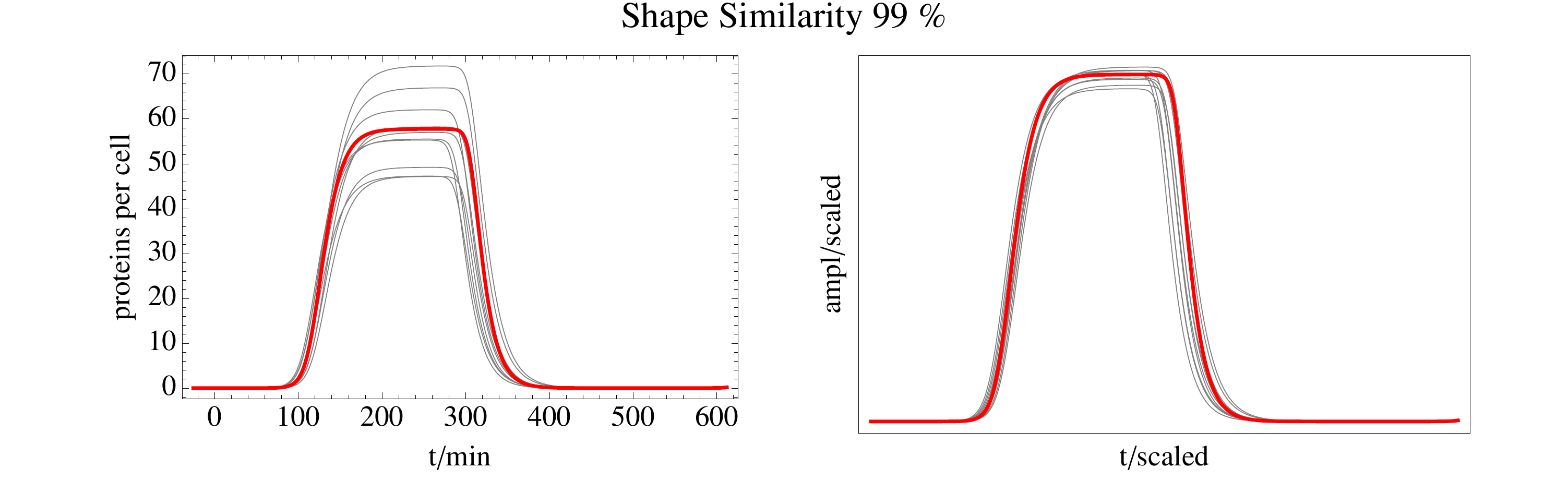}
      \caption{Shape variations under the global parameter changes. On the left we see the numerically observed oscillations, which differ moderately in the period length and amplitude. The reference shape for the unperturbed system is shown in red. On the right the time scale (x-Axis) has been scaled with the period and the maximal amplitude for each curve respectively. This demonstrates how the scaling in time and normalized $L_2$ score assesses the shape similarity.}
      \label{shape}
   \end{center}
\end{figure*}

\section{Bifurcation Software AUTO}
AUTO \cite{Doedel} performs systematic variation of specified parameters and assessment of qualitative characteristics of 
ODE system such as fixed points, periodic orbits and their stability. The ODEs  are of the form:
\begin{equation}
u'(t)=f(u(t),p), \hspace{0.2cm} f(\cdot,\cdot), u(\cdot)\in R^n
\end{equation}
The limited bifurcation analysis is done by solving algebraic equations
\begin{equation}
f(u,p)=0, \hspace{0.2cm}  f(\cdot,\cdot), u\in R^n
\end{equation}
One of the limitations of the package is that the continuation must start nearby guessed/known stable fixed point. 
If such starting point is given, when solution branches are calculated as one or more parameters are gradually changed.

For our system there is a regime, where only one stable fixed point exists (see the Theory section in the main text). 
Starting from that equilibrium point we have performed the continuation gradually changing one parameter and identified
the bifurcation points.

\end{document}